\documentclass[%
reprint,
amsmath,amssymb,
aps,
prd,
]{revtex4-2}
\usepackage{graphicx}
\usepackage{dcolumn}
\usepackage{bm}
\usepackage{hyperref}
\usepackage[mathlines]{lineno}
\usepackage{color}
\usepackage{xcolor}
\usepackage{soul}

\begin{document}
	
	\title{3D magnetorotational supernova simulations of a 20 $M_\odot$ progenitor.}


    
	\author{Ilya A. Kondratyev}
	\email{mrkondratyev95@gmail.com}
	\affiliation{Space Research Institute RAS, Profsoyuznaya street 84/32, Moscow, Russia, 117997}
	\affiliation{National Research University Higher School of Economics, Dept. of Physics, Staraya Basmannaya street 21/4 b.5, Moscow, Russia, 105066}
	
	\author{Sergey G. Moiseenko}
	\email{moiseenko@iki.rssi.ru}
	\affiliation{Space Research Institute RAS, Profsoyuznaya street 84/32, Moscow, Russia, 117997}
	\affiliation{National Research University Higher School of Economics, Dept. of Physics, Staraya Basmannaya street 21/4 b.5, Moscow, Russia, 105066}
	
	\author{Gennady S. Bisnovatyi-Kogan}
	\email{gkogan@cosmos.ru}
	\affiliation{Space Research Institute RAS, Profsoyuznaya street 84/32, Moscow, Russia, 117997}

\begin{abstract}
		
We present three-dimensional (3D) magnetohydrodynamic (MHD) simulations of the collapse and magnetorotational (MR) explosion of a rotating, magnetized $20\,M_\odot$ progenitor, performed using a self-gravitating, ideal-MHD model with an approximate neutrino leakage scheme. We consider three models with weak, medium, and strong initial poloidal magnetic fields. For every model we perform a corresponding two-dimensional (2D) axisymmetric simulation using identical initial conditions. It allows to clarify the effect of dimensionality in the explosion dynamics. For weak- and medium-field models, we find that the magnetic energy in the layer of the protoneutron star, approximately, between 20 and 60 km, undergoes a rapid, exponential growth phase in 3D, with a characteristic growth time of the order of $\tau\sim10$~ms, which is absent in the corresponding 2D calculations. This amplification leads to a more rapid extraction of rotational energy from the PNS compared to the axisymmetric case at the initial stage of MR explosion. Comparing the measured growth timescale to the characteristic timescales of the magnetorotational instability, the Tayler instability, and convection, we find that only the convective timescale is consistent with the observed growth rate.
		
\end{abstract}	
	
\maketitle	
    
\section{Introduction}
\label{sec:introduction}

Gravitational collapse of the core of a massive star and the subsequent supernova explosion (SN) are among the fundamental processes in the evolution of massive stars. In this process the interstellar medium is enriched with heavy elements, and  compact remnants -- neutron stars and stellar-mass black holes -- are formed. 

The first models of core collapsing SNe have been based on neutrino-driven mechanism (see \cite{colgate, bethe1995}, in which the neutrino energy flux from the collapsed core is heating the falling matter, which changes the direction of its motion, what should lead to the SN explosion. For obtaining the energy output, sufficient for explanation of SN, this mechanism had many modifications (see e.g. \cite{blondin2003}), among which the most significant was a discovery of convective instability in the expelled matter \cite{epstein1979}. To date there are lot of papers that are devoted to the neutrino driven mechanism, see e.g. reviews \cite{janka2012, mueller2020}.

For stars with sufficiently rapid core rotation and strong magnetic fields, the magnetorotational (MR) mechanism \cite{BK1970} may play a significant, if not the primary, role in launching the explosion. Interest in this class of models extends beyond the energy-release mechanism itself. The MR scenario naturally links to the formation of rapid neutron-rich bipolar outflows, where r-process nucleosynthesis is possible (see, for example, \cite{winteler2012, halevi2018}), to the origin of strong magnetic fields \cite{thompson1993}, to the central engines of origin of gamma-ray bursts \cite{usov1992} and hypernovae. Observations show (e.g., \cite{utrobin2021}) that some powerful supernovae explode in the form of collimated, jet-like outflows, which could serve as indirect evidence for the viability of this mechanism.

The idea of the MR mechanism was first formulated by Bisnovatyi-Kogan \cite{BK1970}. The first  simplified axisymmetric magnetohydrodynamic (MHD) calculation was done by LeBlanc \& Wilson \cite{lw1970}. The physics of the mechanism is relatively simple: differential rotation of the protoneutron star (PNS) formed in the collapse ``winds up'' the initially poloidal magnetic field, generating a rapidly growing (linearly with time) toroidal component. Once the toroidal field reaches sufficient strength, the magnetic pressure and tension of the field lines become dynamically significant and lead to the formation of an outflow, which effectively removes energy and angular momentum from the inner regions and can drive an explosion even in the absence of sufficient neutrino heating or hydrodynamic mixing. The expelled matter may be collimated along the rotation axis (a magnetic ``tower'' or jet, \cite{lyndenbell1996}).

Calculations of 1D MHD models with simplified physics have been performed in \cite{bk1976, mh1979, ardeljan1979}. They have shown, that a substantial part of the rotational energy is released through the winding of the toroidal field. A systematic study of the MR mechanism became possible only with the transition to 2D axisymmetric models in the early 2000s. The works of various groups \cite{takiwaki2004, ardeljan2005, moiseenko2006, obergaulinger2006, burrows2007} have demonstrated that, with sufficiently rapid rotation and a strong initial magnetic field, the energy output in this model is high enough for SN production.

In the context of 2D MR supernova (MRSN) simulations, also other phenomena have been studied. They include formation of collimated bipolar outflows in magnetically-driven explosion and PNS kicks formation \cite{sawai2008, kondratyev2024}, gravitational-wave signals \cite{takiwaki2011, BK2017, jardine2022}, a black hole/magnetar  formation \cite{obergaulinger2017}. The important feature of this model is development of MHD instabilities as a field-amplification mechanism. In 2D models even a relatively weak initial field is substantially amplified through the magnetorotational instability (MRI) \cite{bh1991, akiyama2003, sawai2014}, or through the axisymmetric, ``sausage-type'' ($m=0$) branch of the Tayler instability \cite{spruit1999, ardeljan2005, moiseenko2006}. It removes a need for unrealistically strong presupernova magnetic fields for obtaining strong explosion.

The transition to three-dimensional calculations (see, e.g., \cite{winteler2012, moesta2014, kuroda2020, bugli2021, powell2023, shankar2026}) became possible relatively recently due to increased computing resources. It had shown a development of non-axisymmetric MHD instabilities, primarily the kink-type instability ($m=1$) \cite{begelman1998, moesta2014} in the toroidally-dominated, jet-like outflow. The development of the kink instability can partially or completely destroy the collimation obtained in axisymmetric calculations, transferring the energy of the ordered field into turbulence, and making the explosion less collimated and less energetic. Others 3D MHD instabilities rapidly amplify the magnetic field to dynamically important strengths in the turbulent region of the PNS. In 3D the MRI develops, through the full-fledged turbulent cascade, contrary to 2D. High-resolution GRMHD simulations, that resolved the fastest-growing MRI wavelength, have shown \cite{moesta2015}, that, in addition to formation   
of a small-scale MRI turbulent structure, an ordered  large-scale mean-field is fully or partially restored, what may support collimated outflow, see also  \cite{reboulsalze2022}.

Aside from the MRI, the evolution of the magnetic field directly in the PNS region can be affected by a number of other instabilities.  Gradients of lepton number and entropy in the PNS may lead to a convective instability. As one of the possible sources of magnetic field amplification in the rotating PNS, a convective dynamo action was proposed  in \cite{thompson1993}. Recent semi-global simulations of convective dynamos \cite{raynaud2020, masada2022} have shown a generation of strong magnetic field  in PNSs. In the papers \cite{mueller2020_1, matsumoto2024, varma2026} it was shown that the  magnetic field growth by convection  (see also \cite{drobyshevski}) can amplify the explosion energy for slowly rotating supernova progenitors.
    
In stably stratified zones of the differentially rotating PNS, however, the wrapped toroidal component of the magnetic field is much stronger than the poloidal one, and the Tayler instability \cite{tayler1973, pitts1985} (with the $m=1$ kink as the most unstable mode) could be a viable mechanism for amplifying the poloidal component there. The Tayler-Spruit (TS) dynamo model based on  the Tayler instability (see the original paper \cite{spruit2002} and a more recent revision \cite{fuller2019}) could explain the field amplification in stably stratified stellar interiors. Recent numerical simulations of stably stratified dynamos in protomagnetars \cite{barrere2025} and in radiative stellar zones \cite{petitdemange2024} show the feasibility of the TS cycle  in PNSs.

Here we perform a series of 3D MHD simulations of the collapse and MR explosion of a rotating $20\,M_\odot$ progenitor using the FOReCAST code \cite{kondratyev2024, kondratyev2026}. We consider three models with different initial poloidal magnetic field strengths, and the same initial angular velocity profile. For each of the three models, we perform a direct comparison with the corresponding 2D axisymmetric simulation, carried out with the same resolution, and identical initial conditions. It allows us to specify the effects, connected with transition from 2D to 3D numerical scheme. 

We find that, for a  "weak" and "medium" initial magnetic field models W-3D and M-3D, see below, the magnetic energy in the PNS undergoes a rapid, exponential growth phase in the 3D models, with a characteristic growth time of order $\sim10$~ms, which is absent in the corresponding 2D calculations. That leads to more rapid extraction of rotational energy from the PNS compared to the axisymmetric case. The comparison with different instabilities has shown that in calculations the e-folding time is inconsistent with both the local MRI growth time and the Alfv{\'e}n/Tayler timescale, while closely matching the convective overturn time. That indicates that convective amplification is the process responsible for the observed field growth and appearance of MHD turbulence.
 
The paper is organized as follows. In Section \ref{sec:setup}, the numerical scheme, and initial conditions for the simulations are reported. In Section \ref{sec:results} we report the results, including the explosion dynamics for 3D and 2D models, evolution of the PNS magnetic field and angular momentum for various initial magnetic field strengths. In the Section \ref{sec:summary} we discuss and summarize the results of the computations. The Appendix shows the results of a simplified convergence study for our 3D numerical simulations.

\section{Numerical setup, and initial models}
\label{sec:setup}

To explore the dynamics of MR explosions, we solve the non-relativistic MHD equations in spherical coordinates $(r,\,\theta,\,\varphi)$, with approximate neutrino cooling/heating source terms and self-gravity. The following system is solved:
\begin{widetext}
\begin{equation}
	\begin{gathered}
	\partial_t \rho + {\nabla}\cdot( \rho\textit{\textbf{v}}) = 0, \\
	\partial_t(\rho v_r) + {\nabla}\cdot \left( \rho v_r \textit{\textbf{v}} - \frac{B_r\textit{\textbf{B}}}{4\pi}\right) + \frac{\partial P_{tot}}{\partial r} = \rho\frac{v_\theta^2+v_\varphi^2}{r}-\frac{B_\theta^2+B_\varphi^2}{4\pi r} - \rho \frac{\partial \Psi}{\partial r} \\ 
    \partial_t (\rho v_\theta) + \nabla \cdot \left(\rho v_\theta \textit{\textbf{v}} - \frac{B_\theta \textit{\textbf{B}}}{4\pi}\right) + \frac{1}{r}\frac{\partial P_{tot}}{\partial \theta} = \rho\frac{ v_\varphi^2\cot\theta - v_rv_\theta}{r} + \frac{B_rB_\theta - B_\varphi^2\cot\theta}{4\pi r} - \frac{\rho}{r}\frac{\partial \Psi}{\partial \theta} \\
	\partial_t(\rho v_\varphi) + {\nabla}\cdot\left( \rho v_\varphi \textit{\textbf{v}} - \frac{B_\varphi\textit{\textbf{B}}}{4\pi}\right) + \frac{1}{r\sin\theta}\frac{\partial P_{tot}}{\partial \varphi} = \frac{B_\theta B_\varphi\cot\theta + B_rB_\varphi}{4\pi r} - \rho\frac{ v_rv_\varphi + v_\theta v_\varphi\cot\theta}{r}- \frac{\rho}{r\sin\theta}\frac{\partial \Psi}{\partial \varphi}, \\
	\partial_t E + {\nabla}\cdot\left((E+P_{tot})\textit{\textbf{v}} - \frac{\textit{\textbf{B}} (\textit{\textbf{B}} \cdot\textit{\textbf{v}})}{4\pi}\right) = -\rho\textit{\textbf{v}}\cdot{\nabla}\Psi + Q_{\nu}, \\
	\partial_t(\rho Y_e) + {\nabla}\cdot ( \rho Y_e\textit{\textbf{v}}) = R_{\nu}, \\
	\partial_t\textit{\textbf{B}} - {\nabla}\times(\textit{\textbf{v}}\times \textit{\textbf{B}}) = 0, \\
	\Delta \Psi = 4\pi G \rho.\\
	\end{gathered}
	\label{MHD}
\end{equation}
\end{widetext}
Here $\rho, P, Y_e$ are the plasma density, pressure and electron fraction, $\textit{\textbf{B}}, \textit{\textbf{v}}$ are the magnetic field and velocity vectors, $E = \rho\frac{\textit{\textbf{v}}^2}{2} + \frac{\textit{\textbf{B}}^2}{8\pi} + \rho e_{int}$ is the total energy density, where $e_{int}$ is the specific internal energy, and $P_{tot} = P + \frac{\textit{\textbf{B}}^2}{8\pi}$ is the total pressure. $\Psi$ is the gravitational potential. The term $R_\nu$ defines the change of $Y_e$ due to weak processes, and $Q_\nu$ is the neutrino source term (cooling or heating) for changes in the thermal energy.

For all models presented here we use the Lattimer-Swesty equation of state (EOS)  \cite{LS1991} with  nuclear incompressibility  $K = 220$ MeV for baryonic matter, together with electrons, positrons, and black-body radiation (see \cite{oconnorott2010} for the details of the resulting EOS table). In simulations of the neutrino processes, we use a ray-by-ray multi-flavor neutrino leakage scheme, according to \cite{rl2003, ruffert1996, oconnorott2010}, where we neglect neutrino pressure.

The terms $R_\nu, Q_\nu$ include emission of neutrinos by matter in beta- (electron flavors only) and thermal/pair processes (all flavors) \cite{rl2003, ruffert1996}. Neutrino absorption is taken into account in the optically thin limit as described in \cite{janka2001, oconnorott2010}. This approach overestimates neutrino heating during the early post-bounce phase. To avoid this artifact, we additionally multiply the heating term by the factor $[1 - \exp(-t_{pb}/t_0)]$, where $t_{pb}$ is the post-bounce time measured in milliseconds, and $t_0 = 120$ ms. Prior to core bounce and the PNS formation, we adopt the approximate deleptonization prescription  \cite{liebendorfer2005}, where instead of solution of the equation for $Y_e$, we used a fitting formula for it.  Self-gravity was computed using the spherical-harmonics expansion method of \cite{mueller1995}. The gravitational potential was modified  by substitution of spherically symmetric general relativistic term instead of the Newtonian one   \cite{marek2006} (Case~B). 

For MHD simulations, we employ our Godunov-type finite-volume code \cite{kondratyev2024, kondratyev2026}. The system \eqref{MHD} is solved in spherical coordinates using the fifth-order $PPM5$ spatial reconstruction \cite{mignone2014} and a third-order explicit Runge-Kutta $RK3$ time-stepping scheme \cite{shu1988}. The Godunov flux is evaluated using the $HLLD$ method \cite{miyoshi2005, minoshima2021}, with a switch to $HLL$ at strong shocks. The induction equation is supplemented with the generalized Lagrange multiplier $\psi$ \cite{dedner2002}, which transports and damps the $\nabla\cdot\textbf{B}$ errors; we refer to \cite{kondratyev2024} for the implementation details and the choice of cleaning speed $c_h$. The numerical resolution for all 3D runs is $N_r = 224,\ N_\theta = 64,\ N_\varphi = 128$, covering the domain $0 \le r \le 3.1\times10^4$ km, $0 \le \theta \le \pi$, $0 \le \varphi \le 2\pi$. The radial grid for $r \le 12$ km is uniform with cell size $\Delta r = 500$ m; for larger radii it is logarithmically stretched. The grid in the angular directions is equally spaced. However, a naive use of a uniform angular grid leads to an excessively restrictive Courant-Friedrichs-Lewy (CFL) condition due to the convergence of cells near the coordinate singularities. To mitigate this, we apply an averaging procedure (the ring-average technique; see \cite{zhang2019, mueller2020}) to the conservative variables, using a limiter-based piecewise-linear reconstruction applied after each Runge-Kutta stage. This technique is similar to the dendritic-grid approach \cite{skinner2019, kondratyev2026}, also implemented in our code, and it gives quantitatively similar results while being simpler and more computationally efficient. In the Appendix we present a minimal resolution study, comparing our model W-3D, where 3D effects are most pronounced, with the same model computed on a grid $\sim1.5$ times finer in all directions in the dynamically relevant post-shock and convective regions. This comparison demonstrates that the presented results are numerically convergent. 

For our MRSN simulations we use progenitor model $s20a28n$ \cite{woosley2007} of a non-rotating, non-magnetized star with a zero-age main-sequence mass $20\,M_\odot$, that has been used in numerous studies of both neutrino-driven and MR explosions (see, e.g., \cite{oconnor2018, aloy2020, kuroda2020, shibagaki2024}). The part of the progenitor  covered by our computational domain has radius $3.1\times 10^4$ km and mass $2.995 M_\odot$. Rotation and a magnetic field are added by hand. We impose a cylindrical rotation profile for the angular velocity of the form (see, e.g., \cite{takiwaki2011, moesta2014}):
\begin{equation}
    \Omega = \Omega_0 \frac{R_0^2}{R^2 + R_0^2}\frac{Z_0^4}{Z^4 + Z_0^4},
    \label{omega}
\end{equation}
where the central angular velocity is $\Omega_0 = 3$~s$^{-1}$, and the parameters controlling the degree of differential rotation are $(R_0, Z_0) = (500, 2000)$~km, with $(R, Z)$ the distances from a given point to the rotation axis and to the equatorial plane, respectively. This profile has a relatively moderate degree of differential rotation, with a precollapse ratio of rotational to gravitational energy of $T/|W| = 0.12\%$. The magnetic field is set via a vector potential of the form (e.g., \cite{suwa2007, moesta2014, matsumoto2020, powell2023}):
\begin{equation}
    A_\varphi = r\frac{B_0}{2}\frac{r_0^3}{r^3 + r_0^3} \sin \theta,
    \label{magfield}
\end{equation}
which defines a uniform, $Z$-aligned magnetic field in regions with $r \ll r_0$ and matches onto a vacuum dipole for $r \gg r_0$. The parameter $r_0 = 1000$~km is fixed for all models, while $B_0$ sets the overall field strength. We construct different MRSN models by varying $B_0$: the weak-field model W-3D has $B_0 = 1/\sqrt{4\pi}\times 10^{12}$~G, the medium-field model M-3D has $B_0 = 10^{12}$~G, and the strong-field model S-3D has $B_0 = \sqrt{4\pi}\times 10^{12}$~G. The initial field strength in the strong-field model coincides with the dipole field adopted in \cite{moesta2014, kuroda2020, bugli2021}. The initial magnetic-to-gravitational energy ratio spans the range from $1.3\times10^{-6}$ to $2.1\times10^{-4}$. In addition, we supplement each three-dimensional model with a two-dimensional axisymmetric simulation, denoted by the suffix ``2D''. The models are listed in Tab.~\ref{tab:table1_models}, together with the numerical resolution in the azimuthal direction.

\begin{table}[h!]
    \begin{center}
		\caption{Labels for the models with different initial magnetic fields $B_0$ (in the units of $10^{12}$~G) and dimensionality (azimuthal grid resolution is given in the last column). For 3D models we use $N_\varphi = 128$, while in axisymmetry $N_\varphi = 1$.}
		\label{tab:table1_models}
		\begin{tabular}{c|c|c}
			\textbf{model} & \textbf{$B_{0,12}$} & \textbf{$N_\varphi$} \\
			\hline
			W-3D & $1/\sqrt{4\pi} \approx 0.282$ & $128$ \\
			M-3D & 1                             & $128$ \\
            S-3D & $\sqrt{4\pi} \approx 3.545$   & $128$ \\
		    W-2D & $1/\sqrt{4\pi} \approx 0.282$ & $1$   \\
			M-2D & 1                             & $1$   \\
			S-2D & $\sqrt{4\pi}\approx 3.545$    & $1$   \\
		\end{tabular}
	\end{center}
\end{table}

\section{Results}
\label{sec:results}

\subsection{Overall dynamics and explosion morphology}
\label{sec:overall_dynamics}

The core collapse depends weakly on the initial magnetic field strength and the dimensionality of the models: the core bounce occurs at $\approx 425$~ms after the start of the simulations. After that, we reset the time counter, and all times are measured from the core bounce moment, $t_{pb}$ in milliseconds. Explosion properties for all six models at the final times of the corresponding calculations  --  the end time $t_{pb, fin}$, the diagnostic explosion energy $E_{expl}$, the ejected mass $M_{ej}$, and the PNS mass $M_{PNS}$ -- are listed in Tab.~\ref{tab:table2_results}. The diagnostic explosion energy and the expelled mass are calculated, using the following expressions:
\begin{equation}
	\begin{gathered}
		E_{expl} = \int_{v_r,E+\rho\Psi>0}(E+\rho\Psi) dV, \\
    	M_{ej} = \int_{v_r,E+\rho\Psi>0}\rho dV
	\end{gathered}
	\label{expldata}
\end{equation}
where $E = \rho\frac{\textit{\textbf{v}}^2}{2} + \frac{\textit{\textbf{B}}^2}{8\pi} + \rho e_{int}$, and integration is performed over the region, where the radial velocity is positive, and the matter has a positive local energy, including the negative  gravitational energy. If a part of the outflowing material exits the computational domain at late post-bounce times, we record its mass and energy fluxes at the domain boundary, integrate them over time, and add to the corresponding quantities inside the computational domain in order to evaluate the expelled mass and explosion energy. The integral quantities in the PNS are calculated over the region, where the density is larger than $10^{11}$ g$\cdot$cm$^{-3}$:
\begin{equation}
	\begin{gathered}
		U_{PNS} = \int_{\rho>10^{11}\,\text{g/cm}^3}U\, dV
	\end{gathered}
	\label{Upns}
\end{equation}
where $U$ is an arbitrary physical quantity; in particular, below we use this expression to evaluate the mass of the PNS, its angular momentum, and magnetic field energy.

\begin{table}[h!]
	\begin{center}
		\caption{ Explosion properties for the simulated models. For each model, the final post-bounce time $t_{pb,fin}$ in milliseconds is presented, followed by the explosion energy $E_{expl}$ (in foe$=10^{51}$~ergs), the expelled mass $M_{ej}$, and the PNS mass (masses are given in the units of Solar mass). }
		\label{tab:table2_results}
		\begin{tabular}{c|c|c|c|c}
			\textbf{model} & \textbf{$t_{pb, fin},$ ms} & \textbf{$E_{expl, 51}$} & \textbf{$M_{ej}/M_\odot$} & \textbf{$M_{PNS}/M_\odot$} \\
			\hline
			W-3D & 1074 & 1.160 & 0.471 & 1.531 \\
			M-3D & 1625 & 1.517 & 0.839 & 1.506 \\
            S-3D & 922  & 3.440 & 0.580 & 1.515 \\
			W-2D & 4500 & 0.862 & 0.385 & 1.943 \\
			M-2D & 1824 & 1.623 & 0.401 & 1.692 \\
			S-2D & 961  & 3.758 & 0.585 & 1.514 \\
		\end{tabular}
	\end{center}
\end{table}

\begin{figure}[!htp]
	\centering
    \includegraphics[width=8.6cm,height=8.7cm]{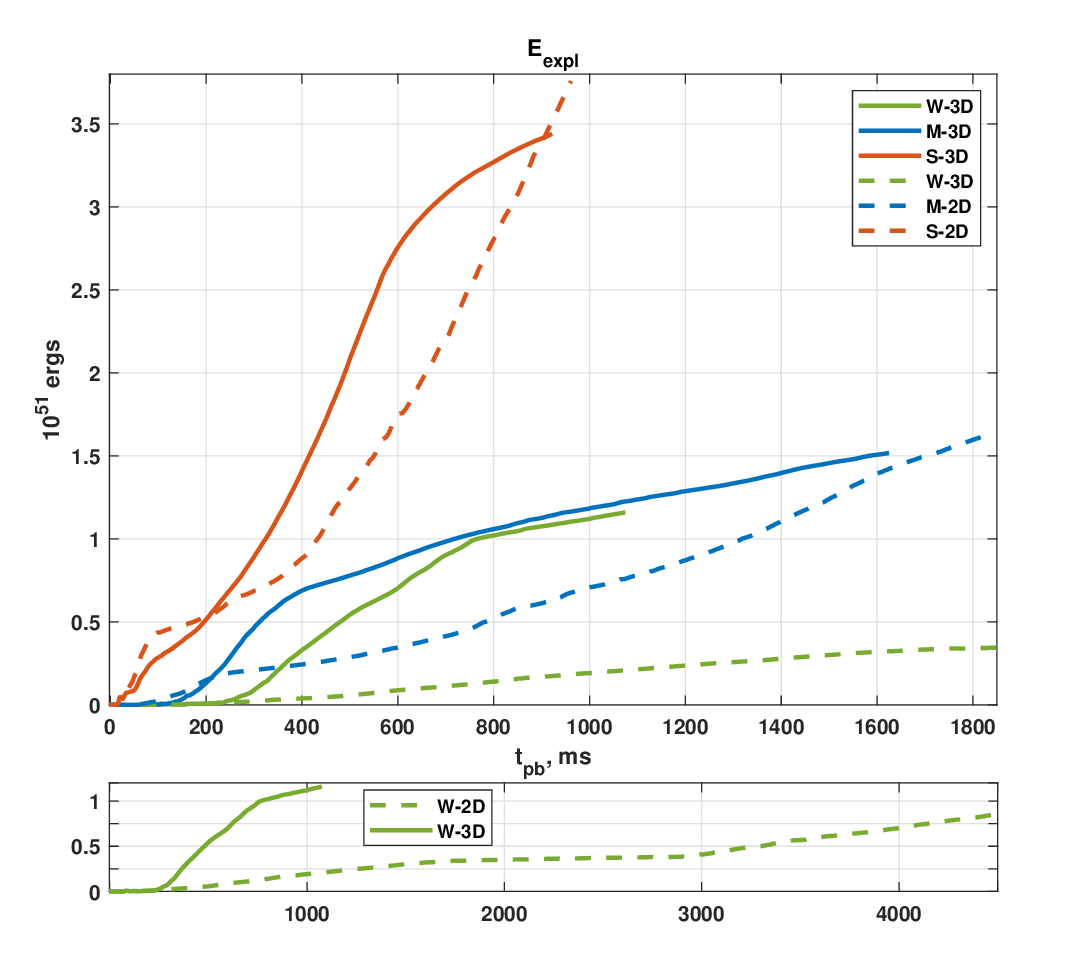}
	\caption{ Evolution of the explosion energy (in units of $10^{51}$~ergs) for the simulated models. Solid lines correspond to the 3D case, while the dashed ones -- to the 2D models. Red lines correspond to the models with strong fields (S), blue lines -- to the medium fields (M), and green ones -- to the weak-field models (W). }
	\label{fig:Eexpl}
\end{figure}

A general picture of the temporal evolution of the explosion energy is presented in Fig.~\ref{fig:Eexpl}. In the two-dimensional simulations the explosion energy grows monotonically without sharp features in all three models, and the order of the curves at any fixed moment of time strictly follows the initial magnitude of the magnetic field: $E_{expl}^{S-2D} > E_{expl}^{M-2D} > E_{expl}^{W-2D}$. At the final post-bounce times in our calculations the explosion energy reaches $\approx 3.8$, $1.6$, and $0.86\times10^{51}$~ergs for the models S-2D, M-2D, and W-2D, respectively. It is worth noting that the explosion energy in 2D is not saturated and still grows at an almost constant rate even after 1 second after the core bounce.

In three-dimensional simulations the final hierarchy by explosion energy ($E_{expl}^{S-3D} > E_{expl}^{M-3D} \gtrsim E_{expl}^{W-3D}$) is preserved in general, the curves demonstrate varying degrees of delay in the energy growth, as well as various subsequent dynamics from model to model, as described below. In 3D models the explosion energies  reach  $\approx 3.44$, $1.52$, and $1.16\times10^{51}$~ergs for  S-3D, M-3D, and W-3D, respectively. The explosion energy in W-3D model is still growing at the final simulation times, and could reach the same values as in M-3D at later times.

Note that in 1D study of MRSN \cite{ardeljan1979}, the time of the explosion depends on the initial value of the magnetic field as $\sim 1/B_0$. Meantime in 1D models the explosion energy does not depend on  initial magnetic field value.

For all three 2D axisymmetric models a relatively standard scenario of MR explosions is observed: differential rotation winds up the initial poloidal field, generating a rapidly growing toroidal component, whose magnetic energy, after some time, can exceed the poloidal one by several orders of magnitude. The resulting Amp\`ere force (given mostly by the toroidal magnetic pressure gradient) becomes dynamically significant and leads to the formation of a collimated, low-$\beta$ jet. Its structure is practically identical for all three models, but the outflow launches the later, the weaker the initial magnetic field. As noted above, at a fixed time, the explosion energy in these models depends significantly on the initial value of the field. Note that for smaller values of initial magnetic fields \cite{kondratyev2024} the dependence of the explosion energy on the  field strength becomes weaker. 

\begin{figure*}[!htp]
	\centering  
    \includegraphics[width=5.8cm,height=4.9cm]{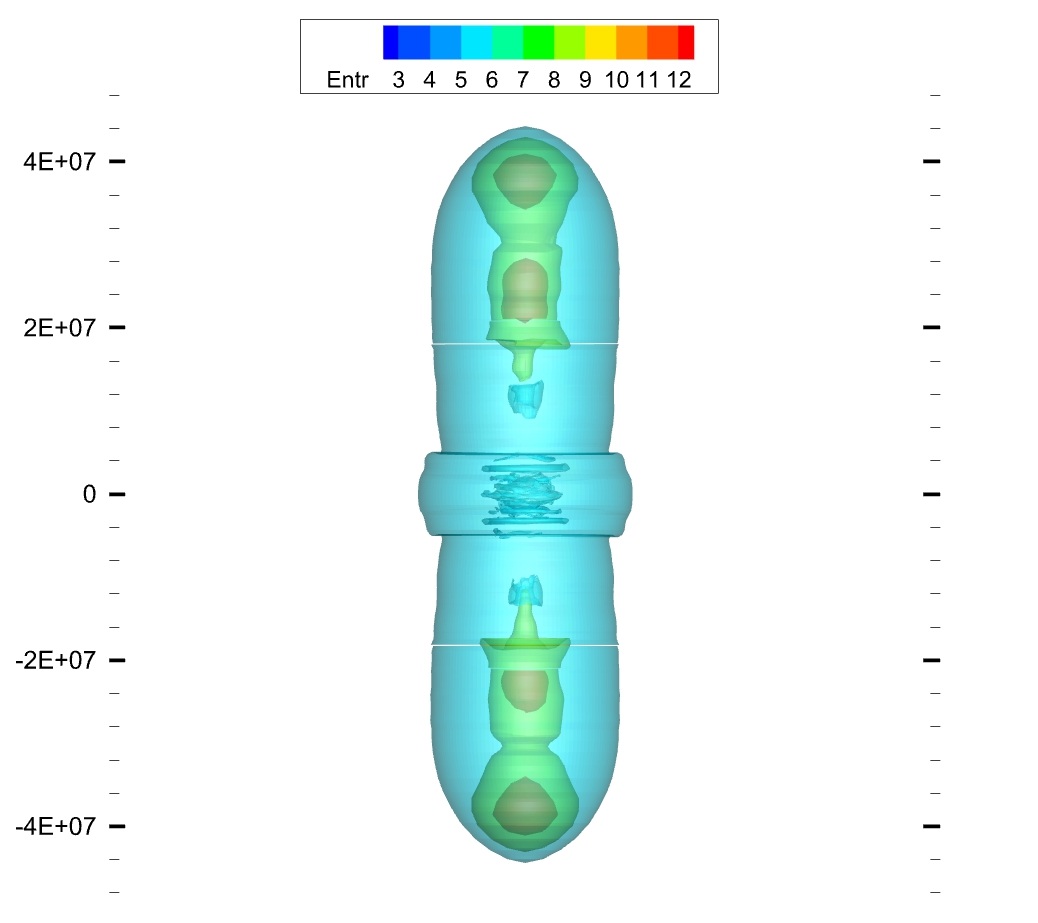}
	\includegraphics[width=5.8cm,height=4.9cm]{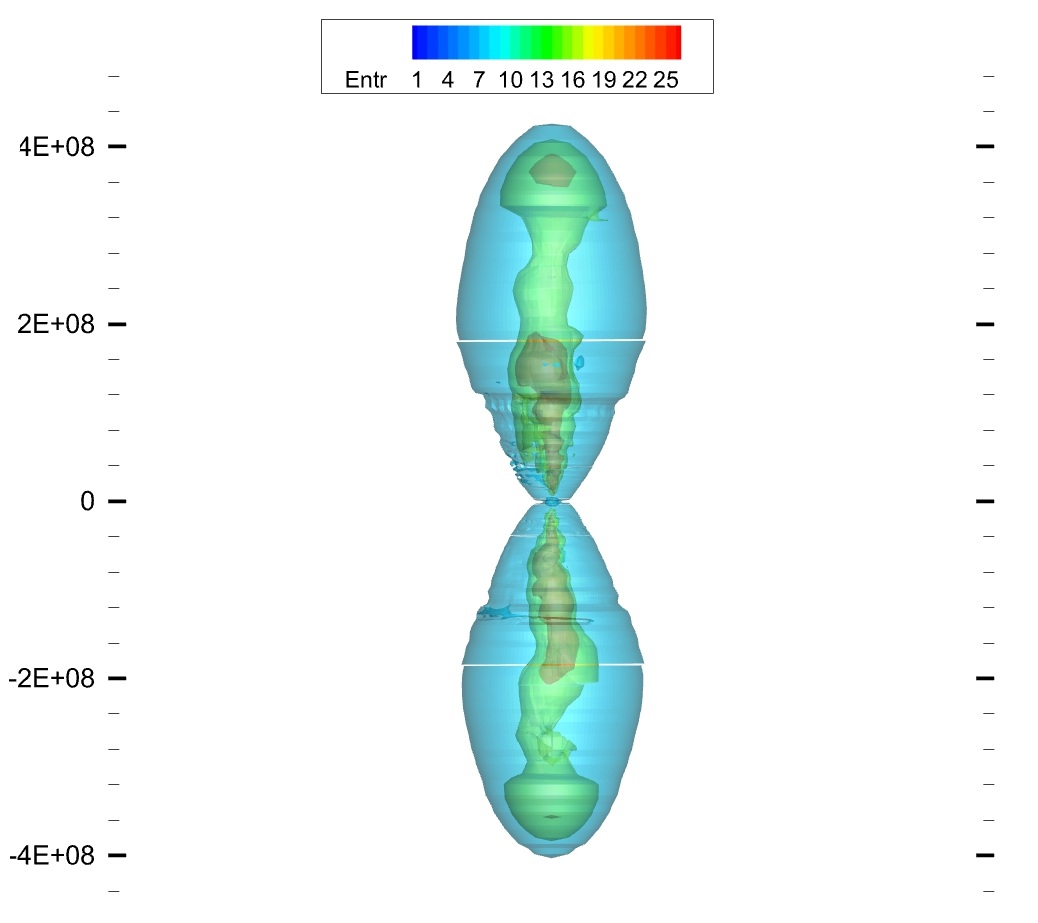}
	\includegraphics[width=5.8cm,height=4.9cm]{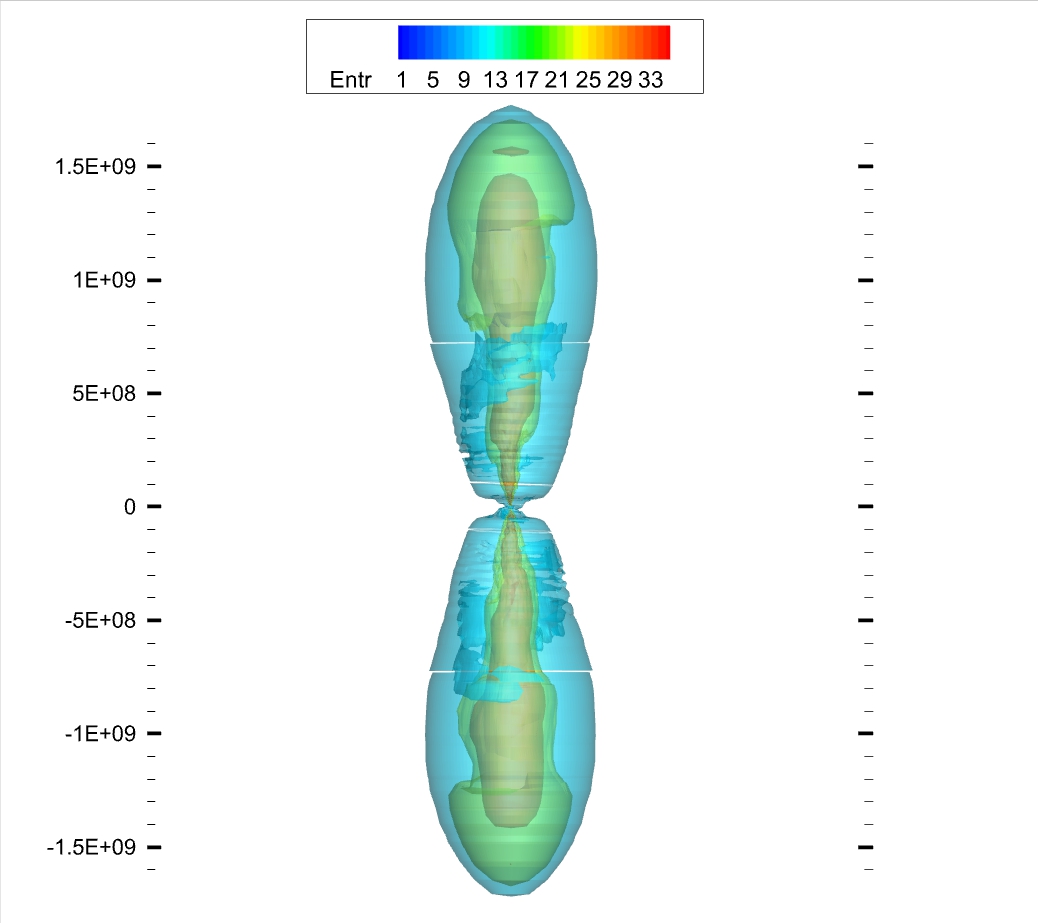}
	\caption{ Volume rendering of the specific entropy (in units $k_B/m_u$) for the strong-field model S-3D at different times: left panel -- $t_{pb} = 31$ ms, central panel -- $t_{pb} = 121$ ms, right panel --  $t_{pb} = 362$ ms. A spatial axes scale is given at the pictures in centimeters. }
	\label{fig:S3D-ent}
\end{figure*}

The dynamics of 3D models is substantially richer. During the first $\sim20$~ms after the core bounce the evolution of all three 3D models is in practice indistinguishable from the corresponding 2D models -- the system evolves almost axisymmetrically. At later times, the evolutionary trajectories of the three models diverge qualitatively, and not just quantitatively.

\begin{figure*}[!htp]
	\centering  
    \includegraphics[width=5.8cm,height=4.9cm]{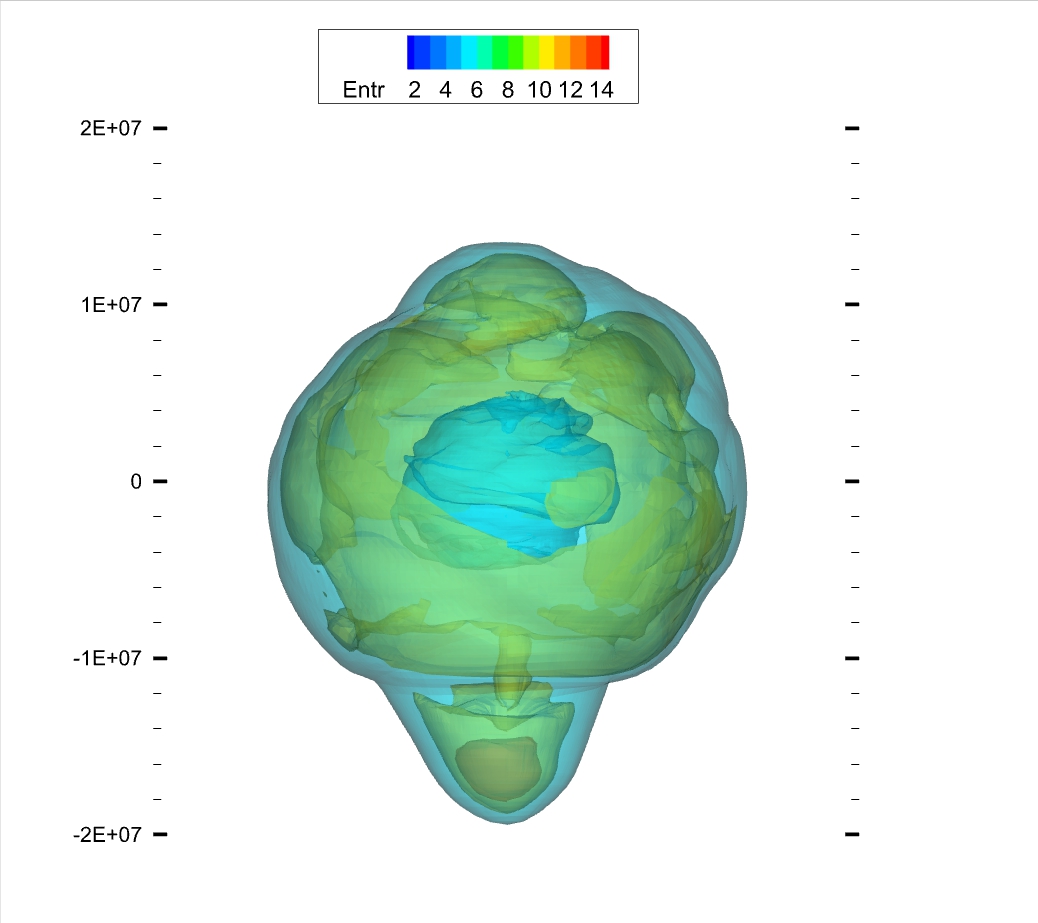}
	\includegraphics[width=5.8cm,height=4.9cm]{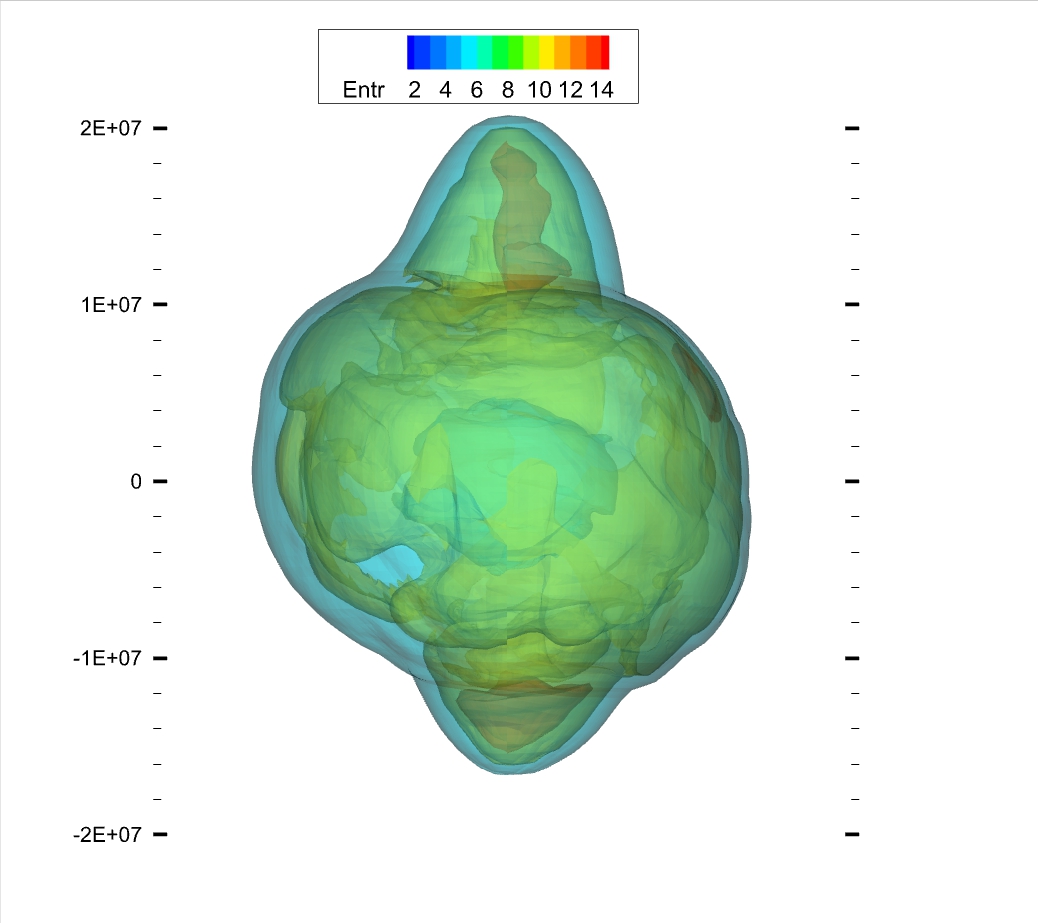}
	\includegraphics[width=5.8cm,height=4.9cm]{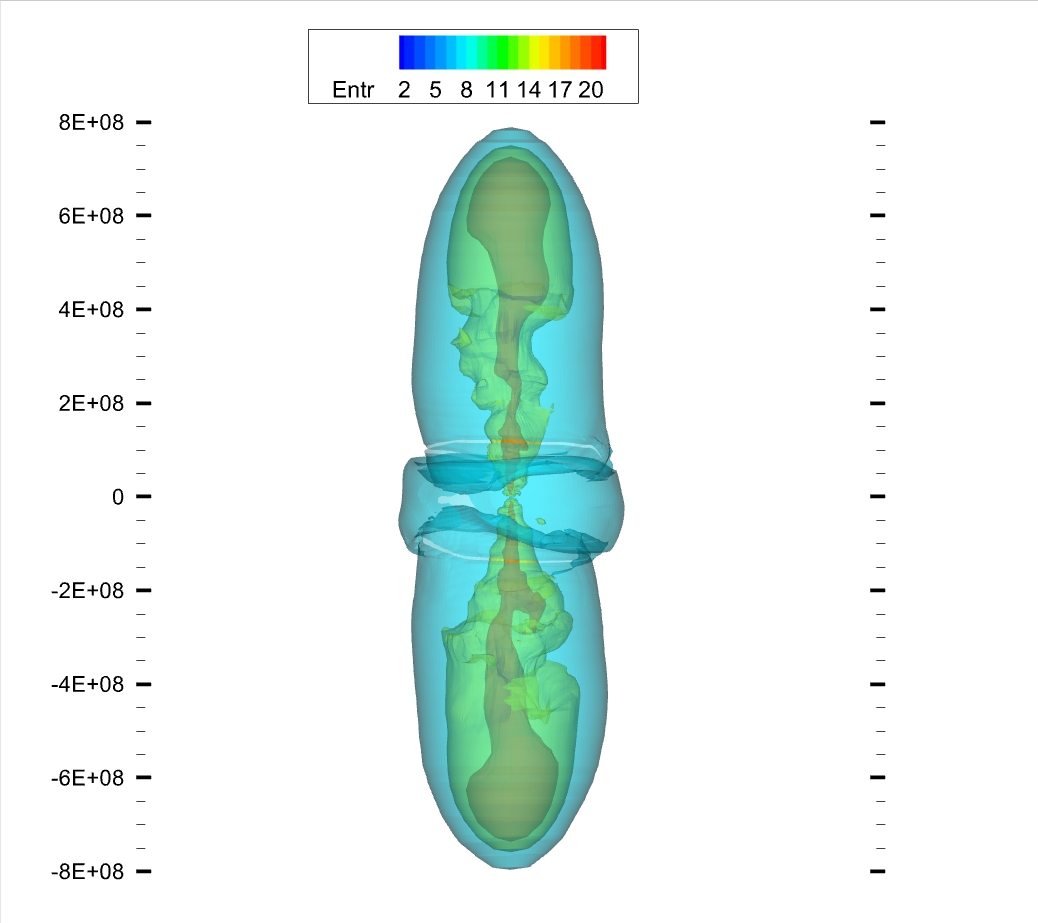}
	\caption{ Volume rendering of the specific entropy (in units $k_B/m_u$) for the medium-field model M-3D at different times: left panel -- $t_{pb} = 73$ ms, central panel -- $t_{pb} = 88$ ms, right panel --  $t_{pb} = 358$ ms. A spatial axes scale is given at the pictures in centimeters. }
	\label{fig:M3D-ent}
\end{figure*}

Snapshots of the specific entropy in the strong-field case S-3D are presented in Fig.~\ref{fig:S3D-ent}. In this model the ejecta form almost promptly ($t_{pb}\lesssim 15$~ms), within the same time as in the corresponding 2D calculation. Three-dimensional effects (fragmentation of scales in the turbulent PNS magnetic field, jet precession) are clearly visible during the subsequent evolution. Their influence on the final integral quantities of the explosion turns out to be moderate. The explosion energy and the ejected mass in the model S-3D are relatively close to the ones for S-2D with a value of $\approx 3.4\times10^{51}$~ergs at the final time moment $t_{pb}=922$~ms. At late times, the growth of the explosion energy starts to saturate, and the corresponding 2D and 3D curves in Fig.~\ref{fig:Eexpl} intersect at $t_{pb}\approx 900$~ms. The results obtained for the strong-field cases can be compared to the findings of Bugli et al. (2021) \cite{bugli2021} (their model L1-0), who considered the same dipole-like field for their 2D/3D MR explosion models, although with a different progenitor and a more sophisticated neutrino treatment. The behavior of the explosion energy for our strong-field models is in a good agreement with the calculations made in \cite{bugli2021} (see their Fig. 2).

In the model with the intermediate magnetic field M-3D, jet formation does not occur immediately. Ejecta, which start to develop at $t_{pb} \sim 70$~ms, stall for some time and finally brake out, subsequently acquiring slight bends and precession. These effects are likely to be produced by the kink instability. Despite this delay in the jet development, the subsequent increase in explosion energy in 3D setting occurs significantly faster than in the corresponding 2D simulation. For later post-bounce times $t_{pb} > 400$~ms the growth of the explosion energy in 3D slows down, while the growth rate in 2D is preserved, leading to a decrease in the differences at the curves for 2D and 3D cases in Fig.~\ref{fig:Eexpl}. It is expected that at $t_{p.b.}\sim 1800$~ms the explosion energy of M-2D will exceed the one in the 3D model, as in the case of the strong initial field. The final ($t_{pb}=1625$~ms) $E_{expl}$ value is $\approx 1.5\times10^{51}$~ergs. To illustrate the described behavior, the specific entropy evolution for the model M-3D is shown in Fig.~\ref{fig:M3D-ent}. 

\begin{figure*}[!htp]
	\centering  
    \includegraphics[width=5.8cm,height=4.9cm]{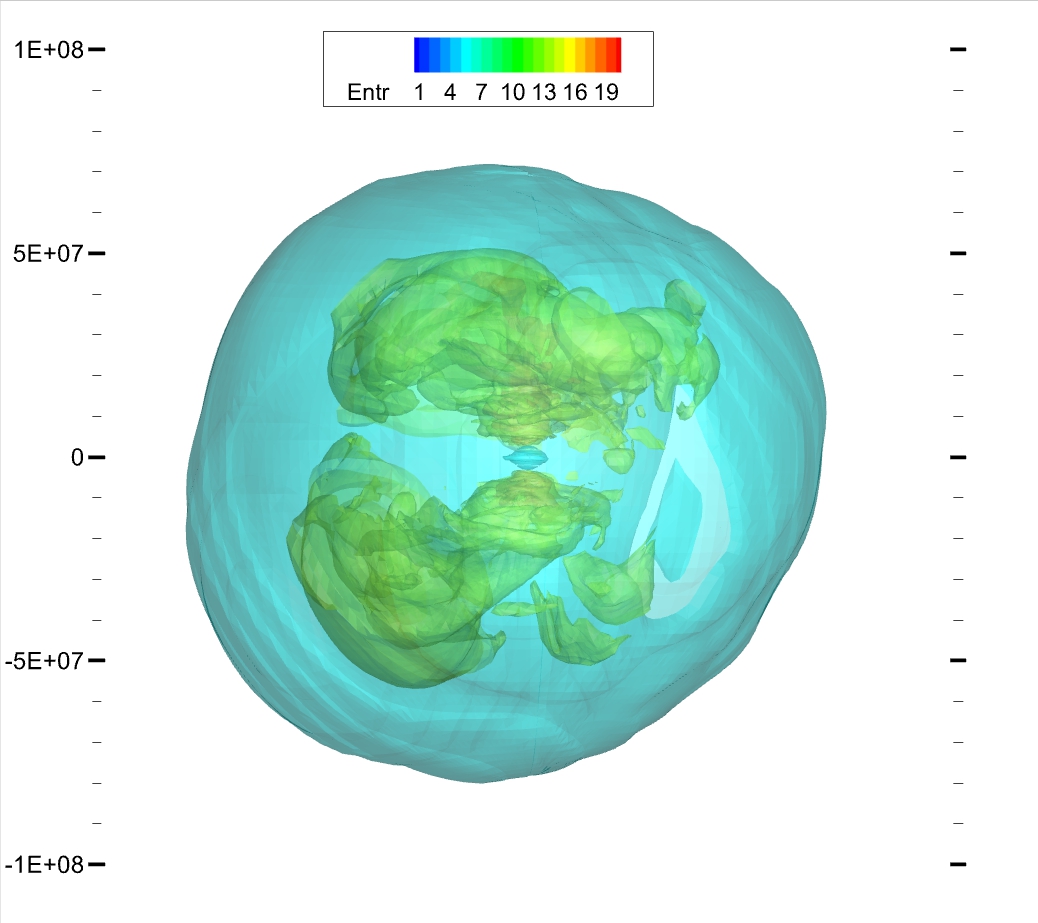}
	\includegraphics[width=5.8cm,height=4.9cm]{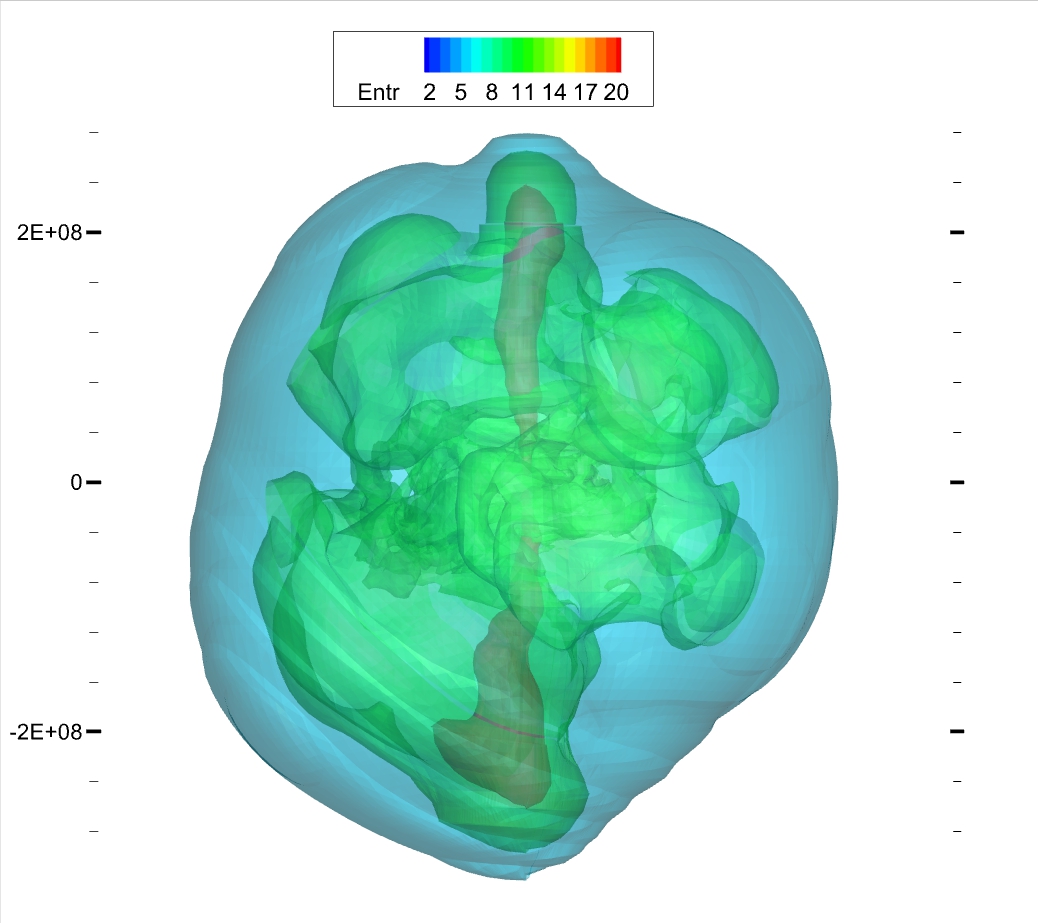}
	\includegraphics[width=5.8cm,height=4.9cm]{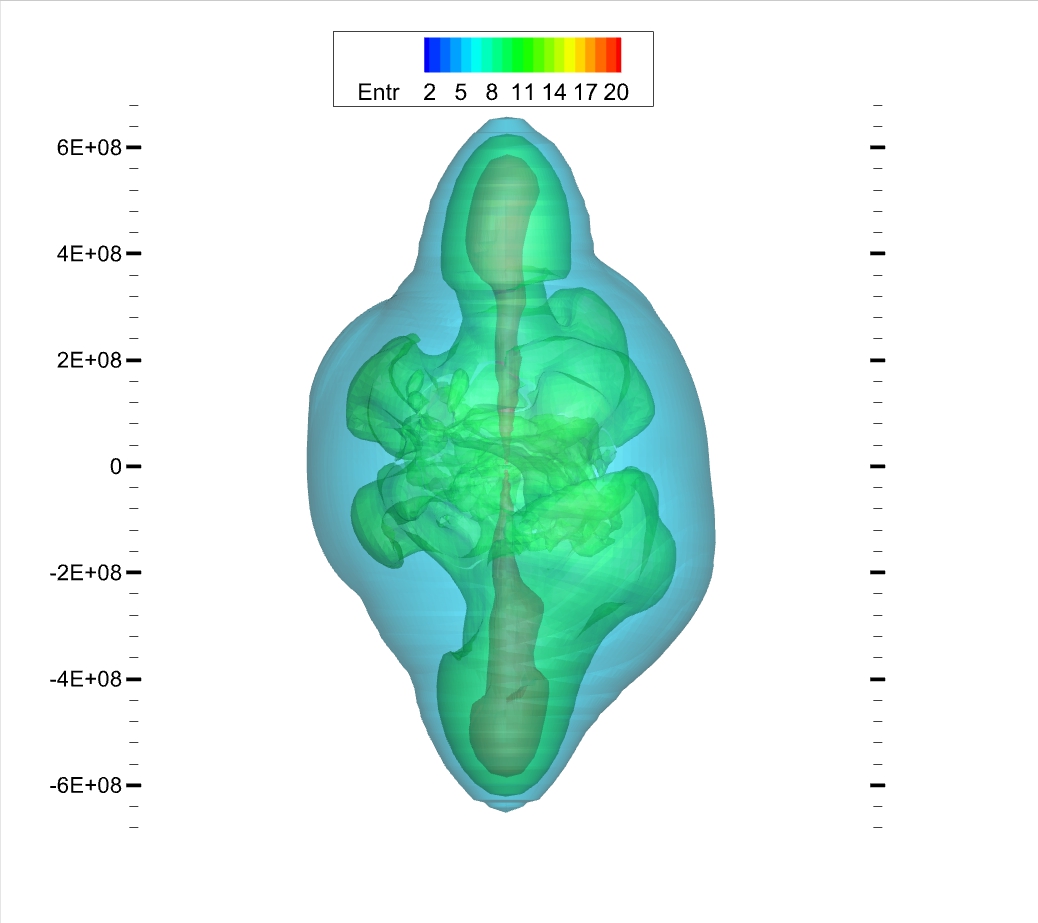}
	\caption{ Volume rendering of the specific entropy (in units $k_B/m_u$) for the weak-field model W-3D at different times: left panel -- $t_{pb} = 313$ ms, central panel -- $t_{pb} = 448$ ms, right panel --  $t_{pb} = 538$ ms. A spatial axes scale is given at the pictures in centimeters. }
	\label{fig:W3D-ent}
\end{figure*}

In the weak-field model W-3D the picture is qualitatively different. The specific entropy evolution for the ejecta in this model is shown in Fig.~\ref{fig:W3D-ent}. Instead of a collimated jet, a slow quasi-spherical expansion of the shell is observed at early times; we associate this with the practically instantaneous destruction of the nascent jet by the kink instability, whose energy is redistributed into the shell. Additionally, the angular momentum is removed predominantly in the equatorial direction by the amplified and disordered magnetic field, feeding the expansion. The collimated outflow begins to form substantially later, after $t_{pb}\sim370$~ms, and also acquires noticeable bends and precession. A somewhat similar evolution of the MR outflows was obtained in 3D simulations of a 39 $M_\odot$ progenitor in \cite{powell2023}. However, in contrast to their findings, we observe that the major fraction of the explosion energy is stored in magnetized jets.

At the end of W-3D calculation ( $t_{pb}\sim 1100$~ms), the difference in the explosion energy for the weak-field model W-3D from W-2D is almost fivefold. It is larger than for the stronger-field models. However, in W-2D model, the explosion energy is far from saturation, while in 3D calculations we observe a considerable decline of angular momentum fluxes, which may correspond to an imminent decrease of MR activity (see below). In order to compare the MR mechanism efficiency in 2D and 3D models for the weak field case, we have extended the calculation of W-2D up to 4.5 seconds after the core bounce. At that times, the explosion energy has reached $\sim 0.86\times 10^{51}$ ergs and is still growing. The final explosion energy in 2D will presumably exceed the one in W-3D at later times, like in the stronger-field models.

The explosion energy in W-3D is similar to the one in model M-3D, so that the energy of the outflows is weakly dependent on the initial field strength in this regime. The final value of the explosion energy (at $ t_{pb}=1100$ ms) is $\approx 1.2\times10^{51}$~ergs. Note that in the models W-3D and M-3D (but not in S-3D) we observe a rapid phase of the poloidal magnetic field  growth in the PNS reaching similar magnitudes. It can explain the similar explosion behavior. The plots and discussion are presented in the next subsection.

Although the moment of the visible explosion launch (the noticeable increase in $E_{expl}$) in the W and M models occurs later in 3D than in the corresponding 2D calculations, the subsequent rate of energy growth in 3D turns out to be higher. Starting from approximately $t_{pb}\sim200$~ms, the explosion energy in the 3D models exceeds those in their axisymmetric analogues. At later times, however, the trend becomes opposite: the energy growth in 3D slows down, while in the 2D models $E_{expl}$ still grows at almost a constant rate. Second, the mass outflow is substantially stronger for the models W-3D and M-3D in comparison to their 2D counterparts, which is clearly noticeable for the intermediate-field models. The mass of the PNS in the W-3D and M-3D models at the final simulation time moments, by contrast, turns out to be approximately $0.15-0.2\,M_\odot$ lower than in the corresponding W-2D and M-2D models.
The strong-field models, although they evolve differently after the bounce in 2D and 3D, end up with similar ejecta properties and the PNS masses at the final times.

\subsection{Post-bounce evolution of the PNS magnetic field}
\label{sec:PNS_mag_field}

\begin{figure*}
\centering
\includegraphics[width=0.48\textwidth]{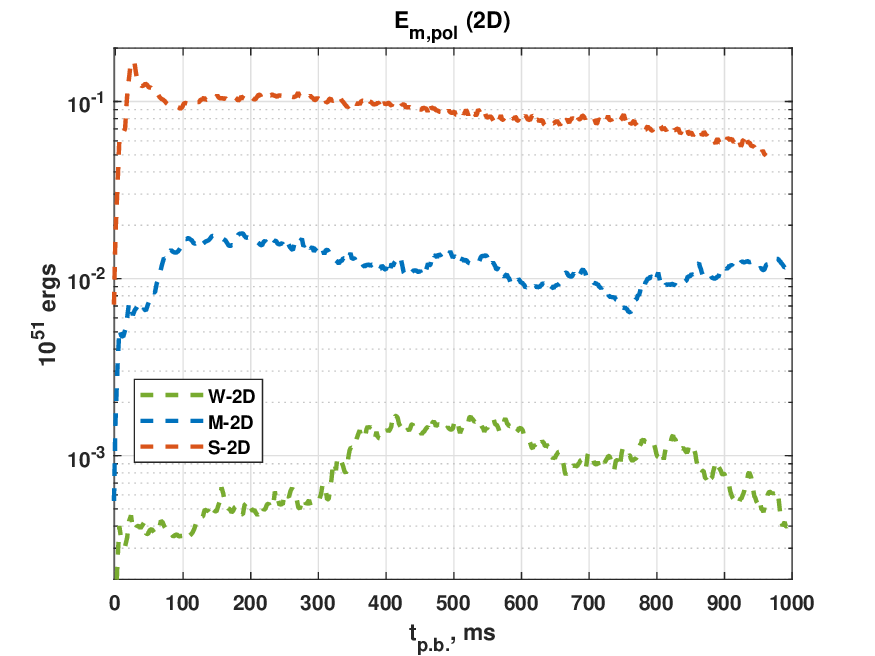}\hfill
\includegraphics[width=0.48\textwidth]{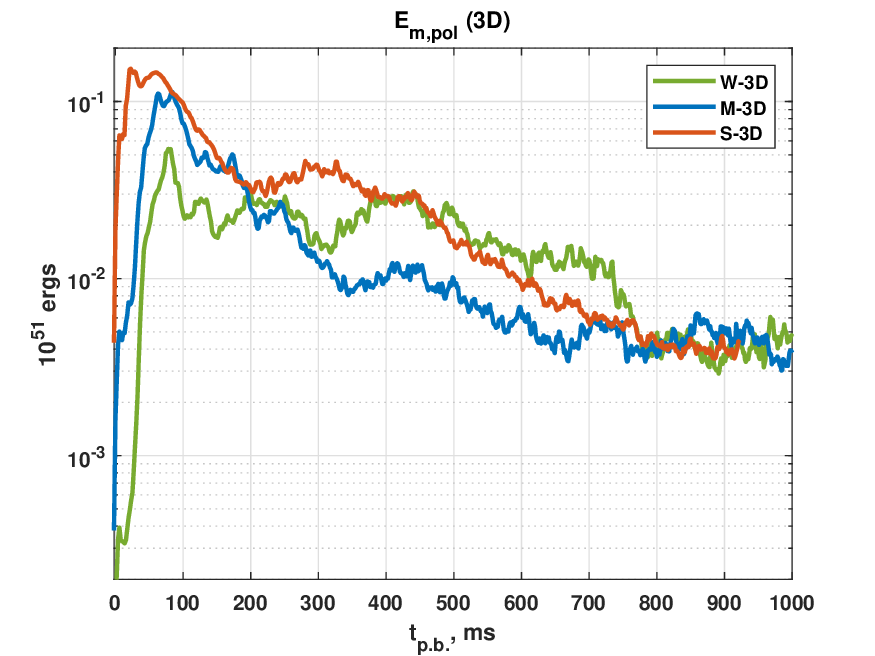}\\[1em]
\includegraphics[width=0.48\textwidth]{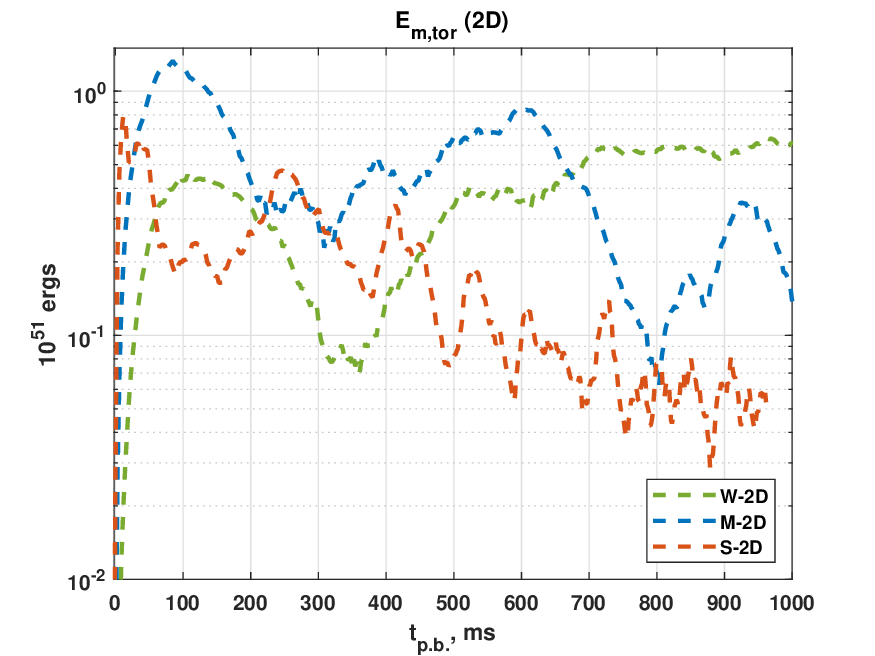}\hfill
\includegraphics[width=0.48\textwidth]{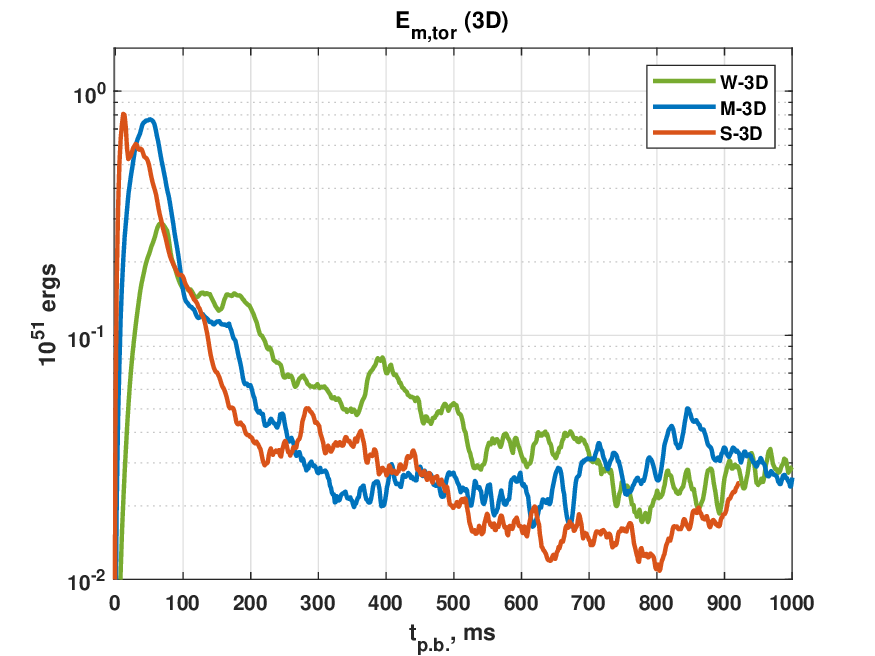}
\caption{ Evolution of the poloidal (top) and toroidal (bottom) magnetic energy inside the PNS for the 2D (left) and 3D (right) models. In the 3D panels, solid lines show the field energy. In the 2D panels, dashed lines show the corresponding axisymmetric values, following the convention of Fig.~\ref{fig:Eexpl}. Colors denote the initial field strength: green -- W, blue -- M, red -- S.}
\label{fig:emag_pns}
\end{figure*}

The evolution of the poloidal and toroidal magnetic energies in the PNS is shown in Fig.~\ref{fig:emag_pns}, with the 2D models in the left panels, and the 3D models in the right ones. After the rapid growth at the bounce, the field behavior in the 2D models does not have any sharp features. The poloidal magnetic energy  tends to roughly preserve its  value. The poloidal field magnitude hierarchy (S-2D $>$ M-2D $>$ W-2D) is strictly maintained over the entire time interval after the bounce. 

The toroidal energy in the 2D models exhibits noticeable oscillations with amplitudes comparable to the toroidal energy itself. The decreases in toroidal energy could be connected to magnetized material being carried away in the outflows, while the subsequent increases are likely due to the accretion of angular-momentum-bearing material onto the PNS, which results in more effective poloidal field wrapping.

At final simulation times, the ratio between the toroidal and poloidal magnetic energies is $E_{m,tor}/E_{m,pol} \gg 1$ for W-2D and M-2D. In contrast to weaker fields, the ratio $E_{m,tor}/E_{m,pol} \sim 1$ for the strong-field model S-2D, likely because a substantial part of the differential rotation was converted into field generation during the first $\sim 500$ ms after the bounce, resulting in a less efficient field wrapping at later times.

\begin{figure*}[!htp]
	\centering  
    \includegraphics[width=5.8cm,height=4.9cm]{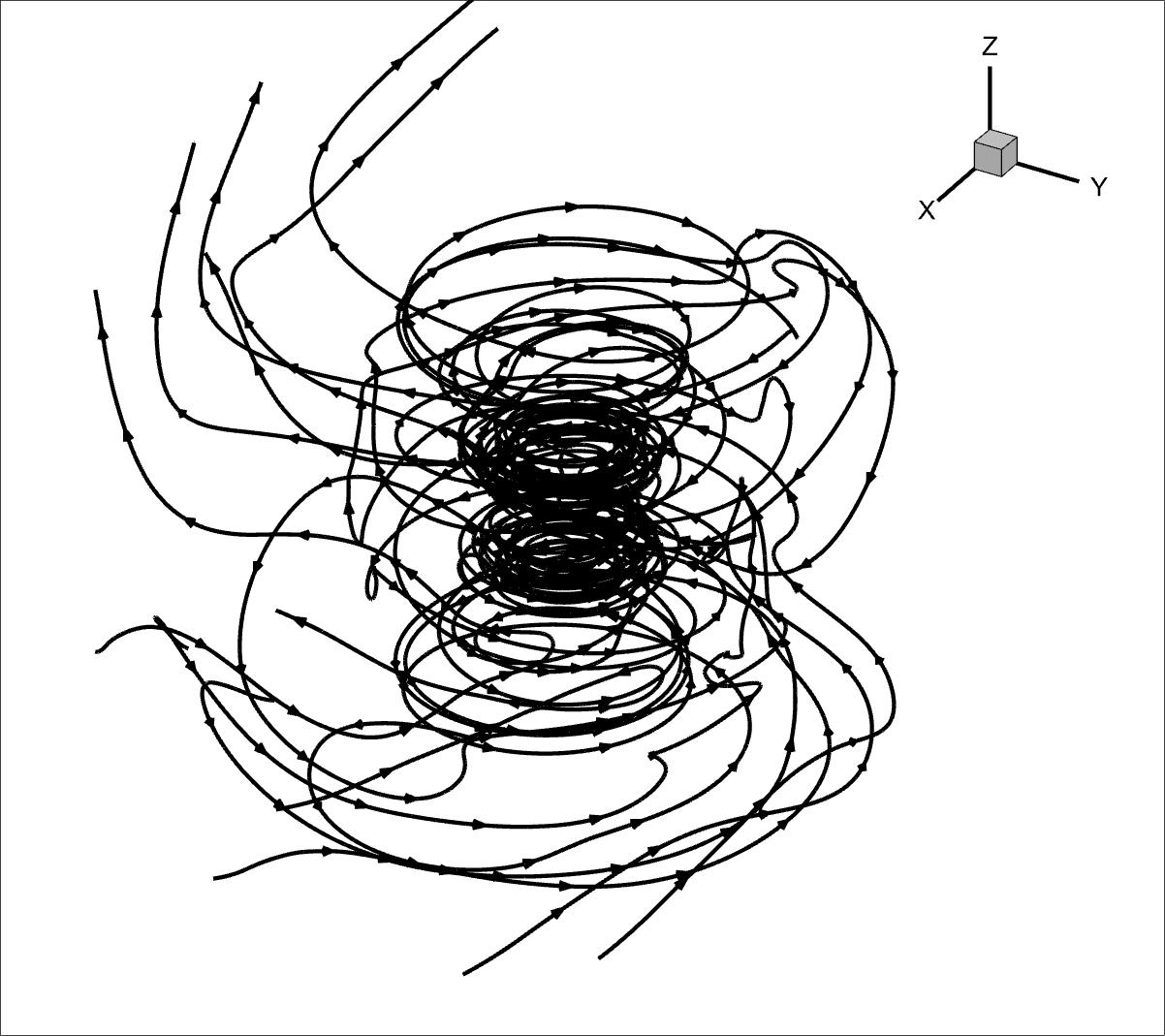}
	\includegraphics[width=5.8cm,height=4.9cm]{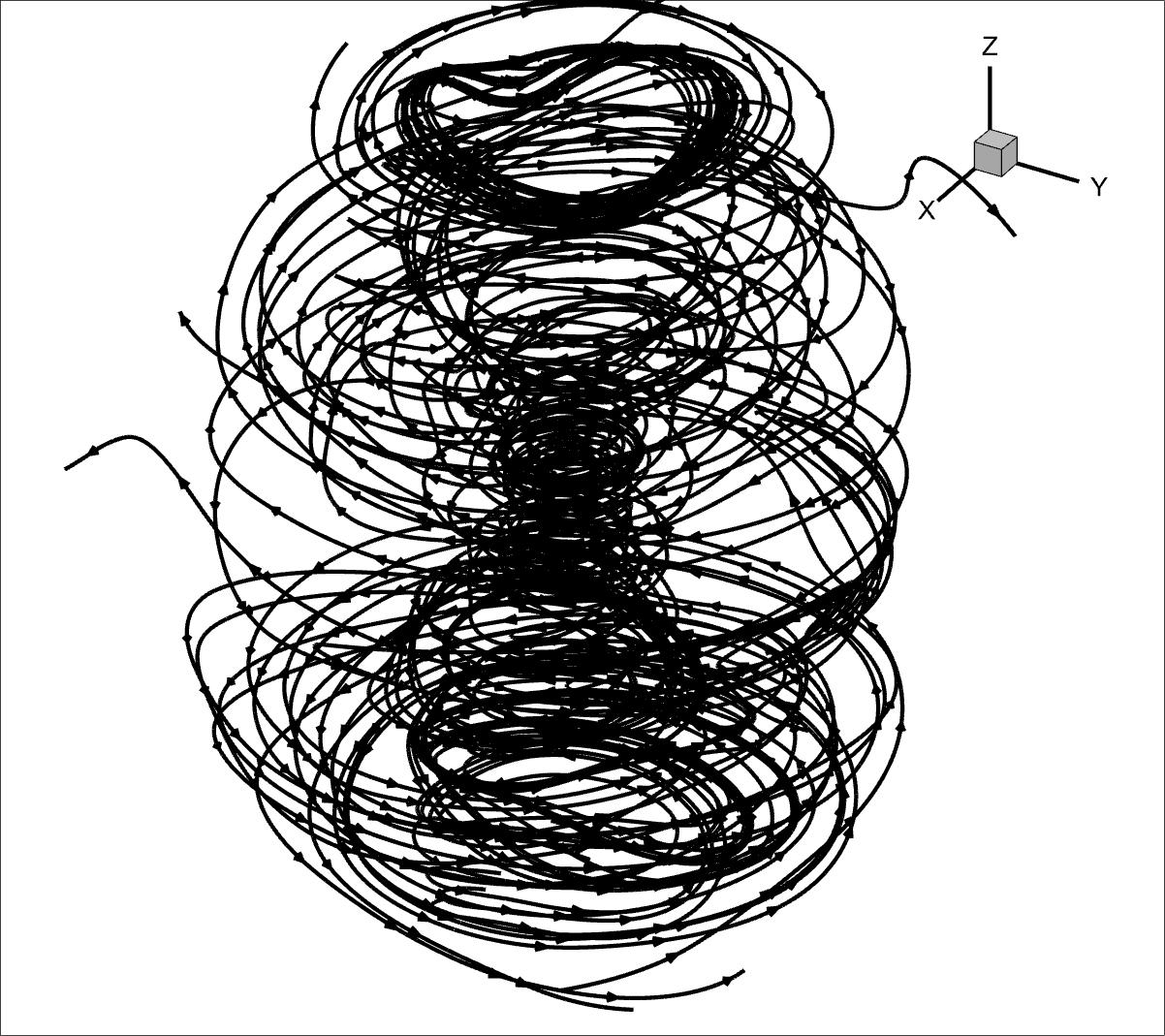}
	\includegraphics[width=5.8cm,height=4.9cm]{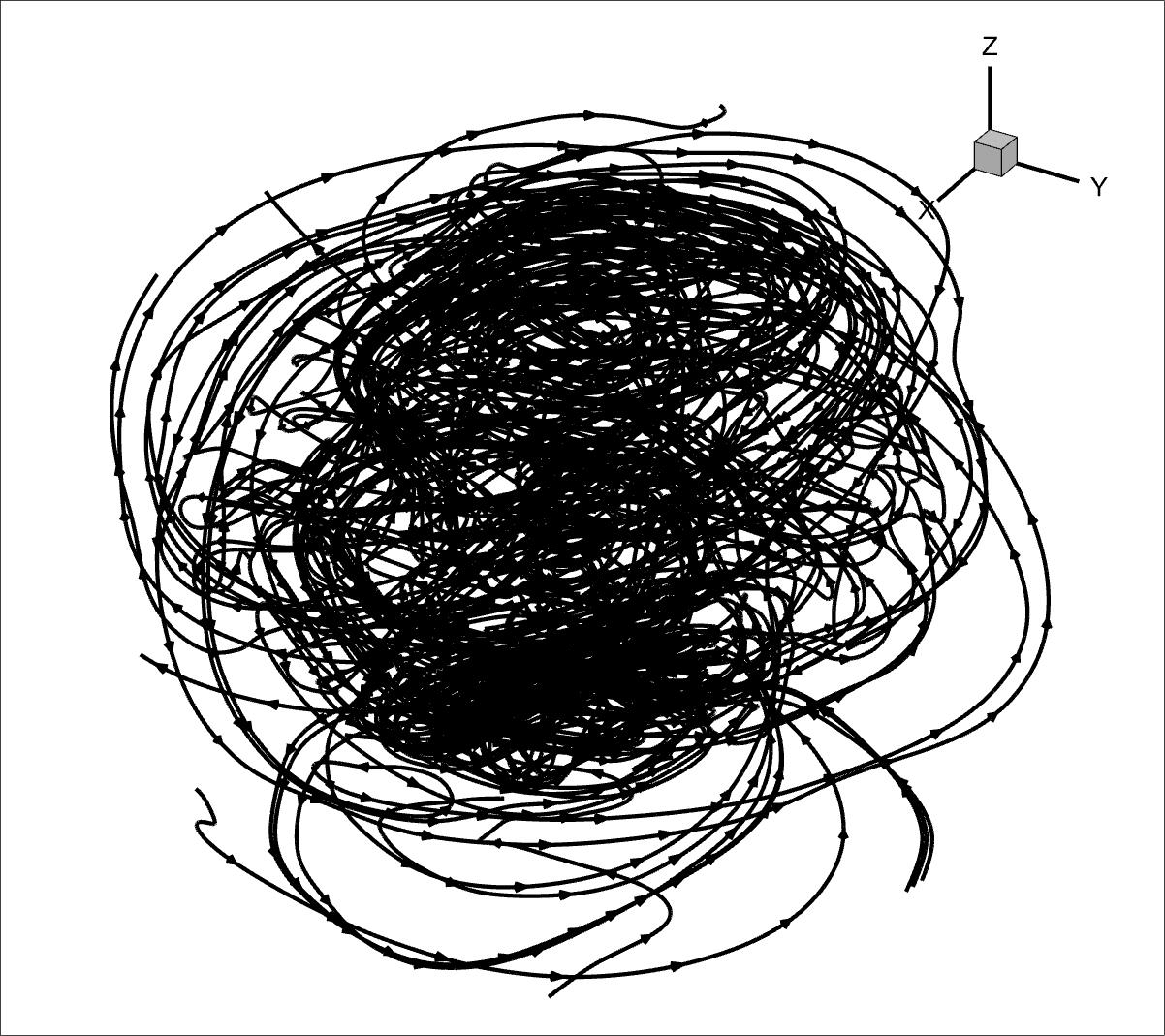}
	\caption{ Illustration of the field topology evolution in vicinity of the PNS during the growth phase in model M-3D. The snapshots of the magnetic field lines are shown at different post-bounce times: left panel -- 12 ms, central panel -- 42 ms, right panel --  72 ms.}
	\label{fig:M3D_field}
\end{figure*}

\begin{figure*}
	\centering
	\includegraphics[width=0.48\textwidth]{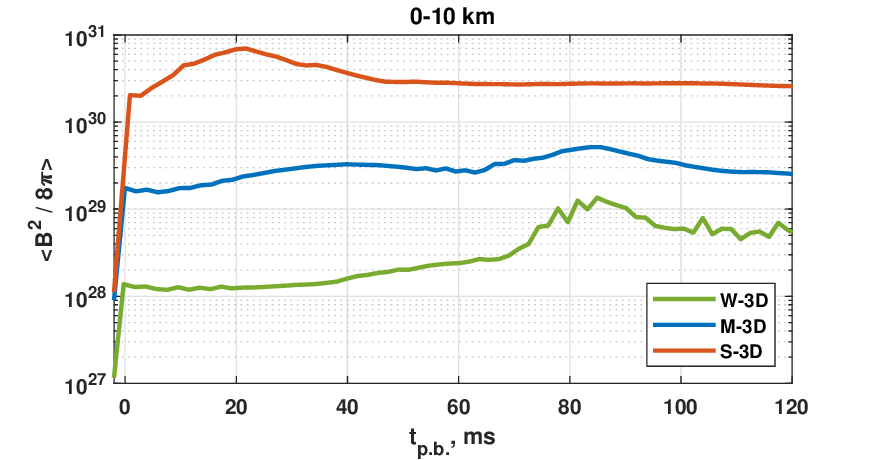}\hfill
	\includegraphics[width=0.48\textwidth]{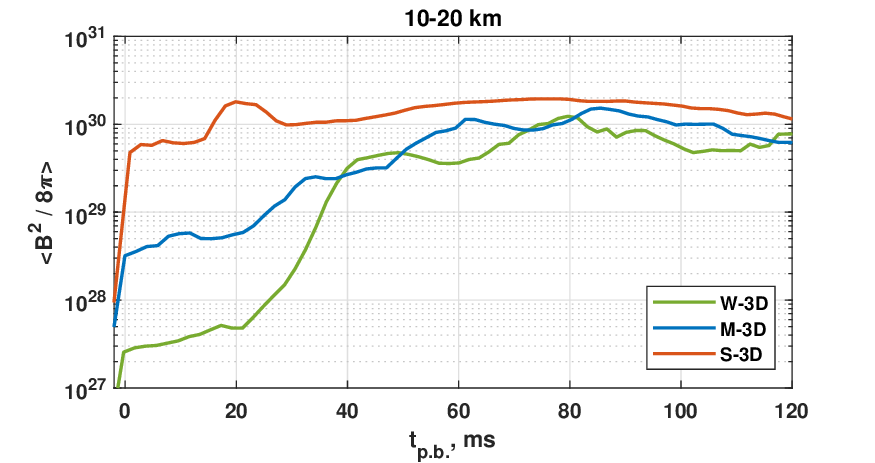}\\[1em]
	\includegraphics[width=0.48\textwidth]{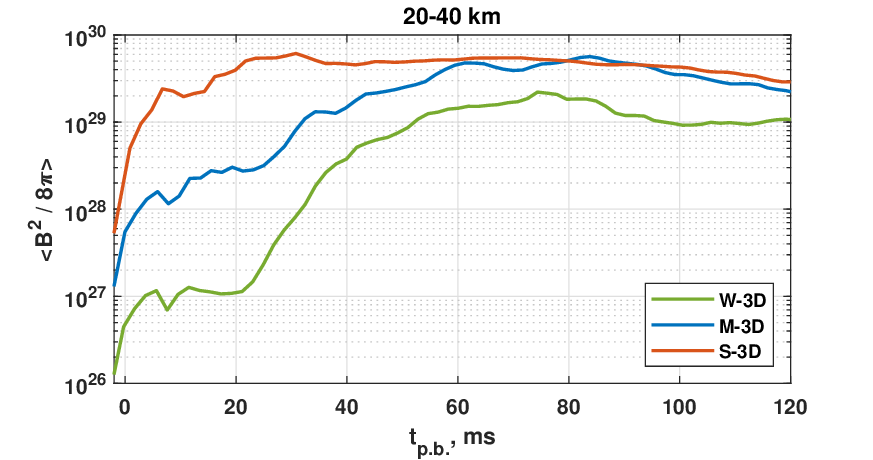}\hfill
	\includegraphics[width=0.48\textwidth]{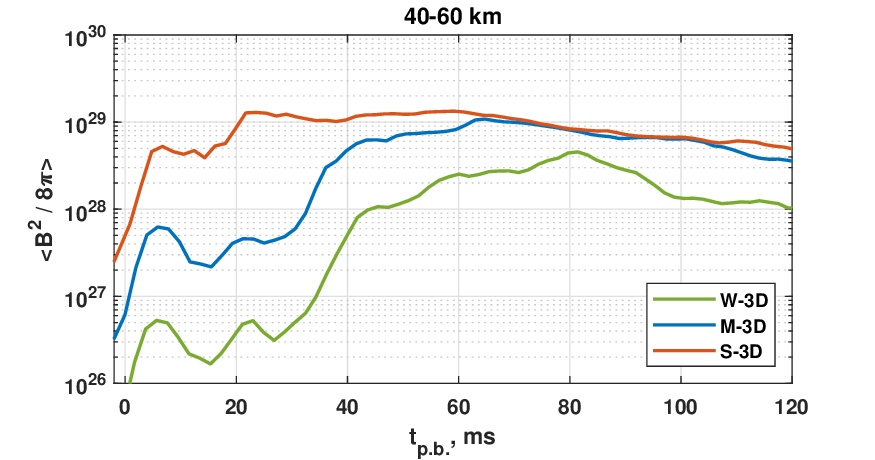}
	\caption{ Volume-averaged poloidal magnetic energy density, $\langle B_{pol}^2\rangle/8\pi$, within four radial shells inside the PNS -- $0-10$, $10-20$, $20-40$, and $40-60$~km -- during the early field-growth phase, for the three 3D models. Colors denote the initial field strength: green -- W, blue -- M, red -- S. }
	\label{fig:emag_shell}
\end{figure*}

The evolution of the poloidal and toroidal magnetic energies in the vicinity of the PNS for three 3D models is shown at the right panels of Fig. ~\ref{fig:emag_pns}. During the first $\sim20$~ms after the bounce, the evolution of all three models is practically axisymmetric, similar to the corresponding 2D calculations. Starting from approximately $20$~ms and up to $55$~ms, the W-3D and M-3D models exhibit a phase of rapid growth of the poloidal field component, which was not observed in our 2D models. The measured e-folding time for the field growth is $t_{e-fold} \sim 10$, and $\sim20$ ms for the models W-3D and M-3D, respectively.
  
Over the same time window, the growth of the toroidal energy for these models slows down in comparison to the 2D case, meaning that some part of the toroidal energy is spent    generating the poloidal component. It saturates at relatively similar values for W-3D and M-3D by the end of this phase. The saturated values of the poloidal energy  have the same order of magnitude in the model S-3D .   
  
After $t_{pb}>70$~ms the nonaxisymmetric component dominates in the poloidal field. At later post-bounce times ($t_{pb}>500$~ms), the ratio between the toroidal and poloidal energies is reduced to  $E_{m,tor}/E_{m,pol} \sim 7$, so that the corresponding field components converge to the same order of magnitude for all models in contrast to axisymmetric simulations.

The illustration of the field growth phase is depicted in Fig.~\ref{fig:M3D_field}. At the very beginning ($t_{pb}=12$~ms) of the post-bounce evolution, the magnetic field has a twisted-torus configuration with $B_{tor}/B_{pol}\gg1$. At later times ($t_{pb}=42$~ms), when the turbulent motion has started, the toroidal loops begin to bend and tilt, leading to the growth of the poloidal component. After the field growth phase ($t_{pb}=72$~ms), the magnetic field acquires a highly non-axisymmetric turbulent structure. 

The evolution of the poloidal magnetic energy density, averaged over the volume of successive spherical shells inside the PNS, is shown in Fig.~\ref{fig:emag_shell} for three 3D models. The final values of the magnetic field in all layers except the innermost shell reach a comparable level (with only the field energy in W-3D being $\sim 2$ times smaller than in the other models), although the path to this level differs between the models. In the innermost layer of the PNS (0-10 km), the field evolves slowly in all cases, and its structure and strength are determined mostly by the compressed progenitor magnetic field. In the shell within 10-20 km, the field grows at significantly different rates. The slope of the S-3D curve on a logarithmic scale decreases as it grows and reaches a plateau most quickly, while M-3D and, especially, W-3D continue to grow longer and gradually approach the S-3D level toward the end of the interval under consideration. It means that the growth rate has an inverse dependence on the initial field strength. In the outer shells (20-40 and 40-60 km), the growth rates of M-3D and W-3D are almost identical after 20 ms p.b. -- after the onset of growth, both curves increase with a similar slope ($\sim (10$~ms$)^{-1}$), differing primarily in the onset of this growth. Model S-3D in both shells reaches saturation earlier with a small increment, that converges towards zero. This behavior is consistent with the fact that in all three models, the field growth saturates upon reaching a comparable energy level, determined by the local properties of turbulence in this region, and not by the initial magnetic field strength per se. Since the initial field in the S-3D model was already closer to this level, saturation occurs earlier for it, whereas M-3D and W-3D approach the same level later, after a longer growth phase.

To determine the mechanism responsible for the observed behavior of the magnetic field in the 3D models, we compare the measured growth rate with the characteristic rates of three candidate instabilities, namely the MRI \cite{velikhov1959}, Tayler modes, and convection. The MRI requires a negative angular velocity gradient, $d\Omega^2/d\log r<0$, in presence of the poloidal field, and has a growth rate  $\sigma$  independent of the magnetic field strength with $\sigma \sim |\frac{d\Omega}{d\log r}|$ for the fastest-growing mode with the wavelength $\lambda_{FGM}\sim v_A/\Omega$ \cite{akiyama2003, obergaulinger2009}. However, the fastest-growing MRI wavelength is not resolved on the grid in the global simulations reported here; we also do not observe channel-mode structure (e.g., \cite{obergaulinger2009, sawai2014, moesta2015}) characteristic of MRI-driven amplification shortly after the bounce, so that the contribution of the MRI to the observed field growth is considered unlikely. The instability itself may be suppressed by the inclusion of the  neutrino viscosity in the PNS \citep{guilet2015}. 

\begin{figure*}[!htp]
	\centering  
    \includegraphics[width=17.4cm,height=7.8cm]{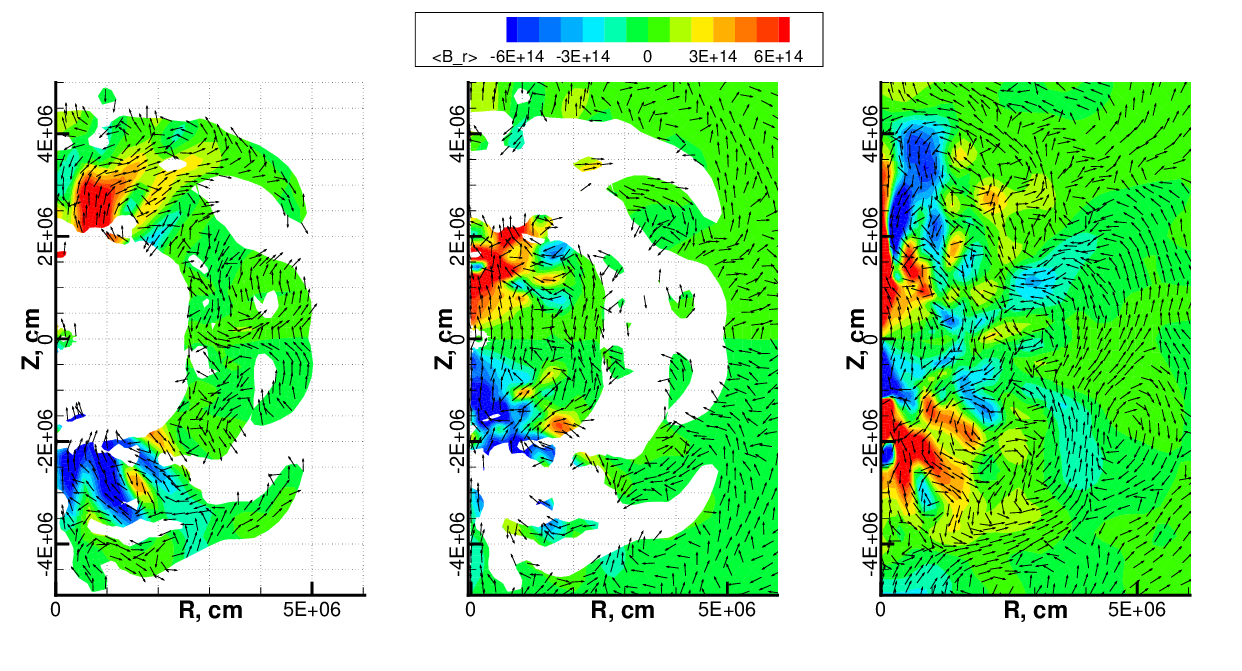}
	\caption{ Azimuthally averaged radial magnetic field $B_r$  behavior (colors represent the induction value. Left panel shows only the convectively unstable zones with $\langle N_{BV}^2\rangle _\varphi \;\leq \;0$ at 30 ms p.b., while central panel shows the stable zones at the same time; Right panel shows the structure of $\langle B_r\rangle_\varphi$ at $t_{pb} = 107$~ms. The squared buoyancy frequency is calculated, using the azimuthally averaged physical quantities. Black arrows show the magnetic field lines of the averaged poloidal component. }
	\label{fig:Bfield_W3D}
\end{figure*}

The Tayler instability ($m=1$) requires a dominant toroidal component of the magnetic field with a sufficiently flat profile ($p=d\log B_\varphi/d\log R>-1/2$) \cite{tayler1973, spruit1999}. In addition, the matter has to be stably stratified with $N_{BV}^2>0$, where $N_{BV}$ is the buoyancy (Brunt-V{\"a}is{\"a}l{\"a}) frequency defined as \cite{kippenhahn2012}

\begin{equation}
	\begin{gathered}
	    N_{BV}^2 = \frac{\nabla \Psi}{\rho}\cdot\left(\frac{\nabla P}{c_s^2} - \nabla\rho\right),
	\end{gathered}
	\label{Nbv2}
\end{equation}
where $c_s$ is the local sound speed. The increment of the Tayler instability depends on the local field strength as $\sigma\approx\omega_{A,\varphi} = B_\varphi/(R\sqrt{4\pi\rho})$ for $\omega_{A,\varphi} \gg \Omega$ and $\sigma\approx\omega_{A,\varphi}^2/\Omega$ for $\omega_{A,\varphi} \ll \Omega$ \cite{pitts1985}. Although the field profile criterion is indeed satisfied near the poles for the weak rotation limit (we have $\omega_{A,\varphi} \lesssim \Omega$ in the regions with intensive field wrapping after the early axisymmetric phase lasting 20-30 ms after the bounce), the growth of the poloidal field in our calculations occurs predominantly in regions with $N_{BV}^2<0$,  i.e. where the matter is unstable, rather than stably stratified, as required for the Tayler criterion. Snapshots of the azimuthally averaged $r$-component of the magnetic field during the exponential growth phase are shown in Fig.~\ref{fig:Bfield_W3D} (left and central panels) for model W-3D. The growth rate measured in our simulations is not proportional to $\omega_{A,\varphi}$ (see Figs.~\ref{fig:emag_pns} and \ref{fig:emag_shell}), which also rules out this instability.

In our simulations, convection starts to operate in the outer layers of the PNS ($20-25$ km $\lesssim r\lesssim 50-55$ km) $\sim 10$ ms after the bounce. For convection driven by a negative lepton-number gradient in the PNS, the instability criterion is likewise $N_{BV}^2<0$ \cite{epstein1979}. The growth rate is estimated as $\sigma\approx\tau_{conv}^{-1}$, where $\tau_{conv}=\Delta_{conv}/v_{conv}$ is the convective overturn time (see, e.g., \cite{masada2022}), with $\Delta_{conv}$ and $v_{conv}$ the characteristic eddy size and mean turbulent velocity in the convection zone, respectively. Based on our measurements of these quantities during the field growth phase, we find $\tau_{conv} \sim(10\text{~km})/(1-1.5 \times  10^8\text{~cm}\cdot\text{s}^{-1})\sim 7-10$~ms, a value in agreement with the measured growth rate.
 
The slow decay of the magnetic field observed in all 3D models is likely to be the result of developed turbulence, which leads to the fragmentation and creation of the small scale magnetic structures (see the right panel in Fig.~\ref{fig:Bfield_W3D}, where the small-scale structures start to form). In numerical models, the smallest scale is the size of the grid cell, and the numerical resistivity could dissipate the small-scale magnetic fields, leading to their damping.

\subsection{Efficiency of MR explosions in 2D and 3D}
\label{sec:explosion_efficiency}

\subsubsection{Angular momentum transport}

\begin{figure}[!htp]
	\centering
    \includegraphics[width=8.6cm,height=5.7cm]{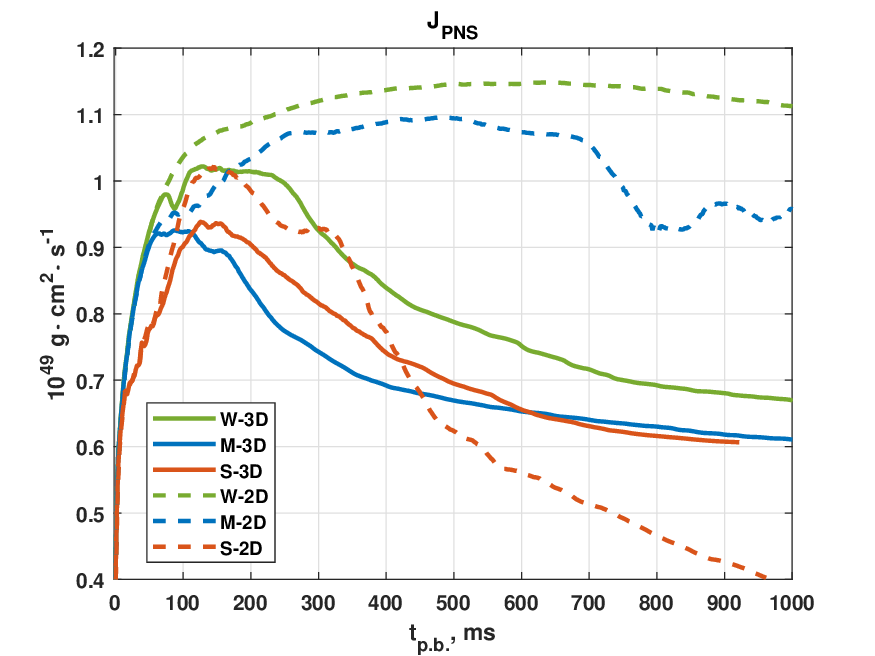}
	\caption{ Evolution of the total PNS angular momentum evaluated with Eq.\eqref{Upns} for all models in the units of $J_{49} = 10^{49} g\cdot cm^2\cdot s^{-1}$. Solid lines correspond to the 3D models, dashed lines -- to the corresponding 2D models. Colors denote the initial field strength: green -- W, blue -- M, red -- S. }
	\label{fig:Jpns}
\end{figure}

The evolution of the total angular momentum contained in the PNS is shown in Fig. ~\ref{fig:Jpns}. In all six models, the angular momentum increases steeply in the first $\sim100$~ms after the bounce due to matter fall-back. The 3D models reach a maximum at $t_{pb}\sim120-150$~ms, with W-3D peaking  higher ($J_{PNS,49} \sim 1.0$) and maintaining a high level longer, M-3D and S-3D peak at $J_{PNS,49} \sim 0.93-0.94$ and begin to decline earlier. Subsequently, all three 3D models monotonically lose angular momentum. All models maintain a comparable rate of angular momentum loss, and M-3D and S-3D practically merge starting from $t_{pb}\sim600$~ms, while W-3D remains higher until the end of the calculations. The PNS angular momentum tends to saturate at a constant value at late post-bounce times. The PNS in 2D models loses its angular momentum differently: W-2D and M-2D reach their maximum significantly later than the 3D models ($t_{pb}\sim 500$~ms) and at a higher level ($J_{PNS,49} \sim 1.10-1.15$) with a slow further decrease. W-2D remains above all other models throughout the entire evolution, while M-2D experiences a noticeable drop at $t_{pb}\sim700-750$~ms, then reaches a roughly constant oscillating level. The drop in $J_{PNS}$ coincides with a decrease in the toroidal field, noticeable in Fig.~\ref{fig:emag_pns}. The S-2D model deviates from the other models most clearly: after reaching its maximum, it loses angular momentum steeply and monotonically. Its curve intersects all 3D models at $t_{pb}\sim400$~ms, ending up below all others by the end of the calculation. Indeed, the strong and ordered magnetic field in this model is able to maintain a continuous magnetic braking. This behavior is consistent with the fact that the explosion energy in the S-2D model, unlike in the 3D case, does not saturate and continues to grow at a nearly constant rate until the end of the calculation (see Section \ref{sec:overall_dynamics}), which requires a continuous effective magnetic torque from the PNS. 

\begin{figure}[!htp]
	\centering
    \includegraphics[width=8.6cm,height=5.7cm]{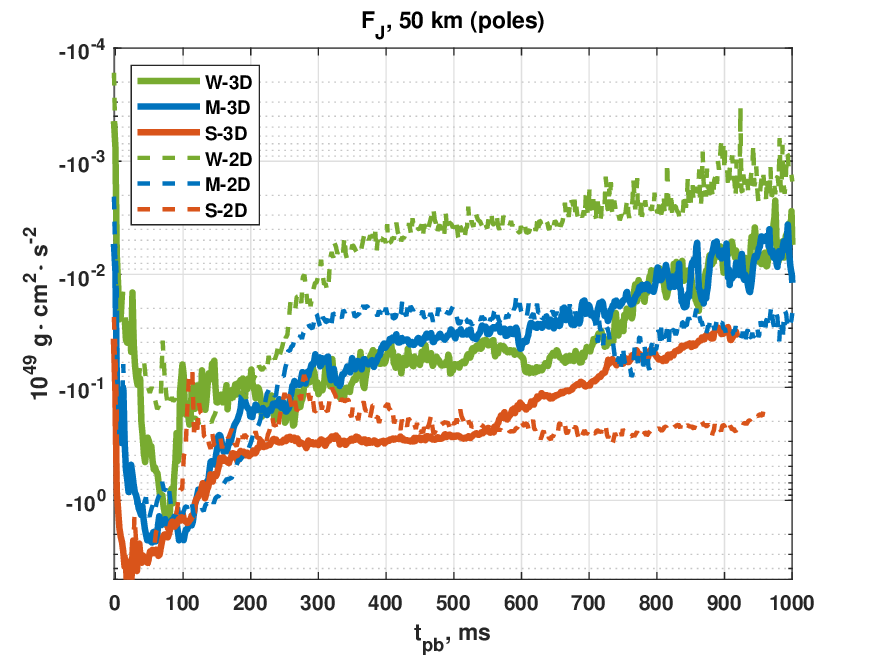}
	\includegraphics[width=8.6cm,height=5.7cm]{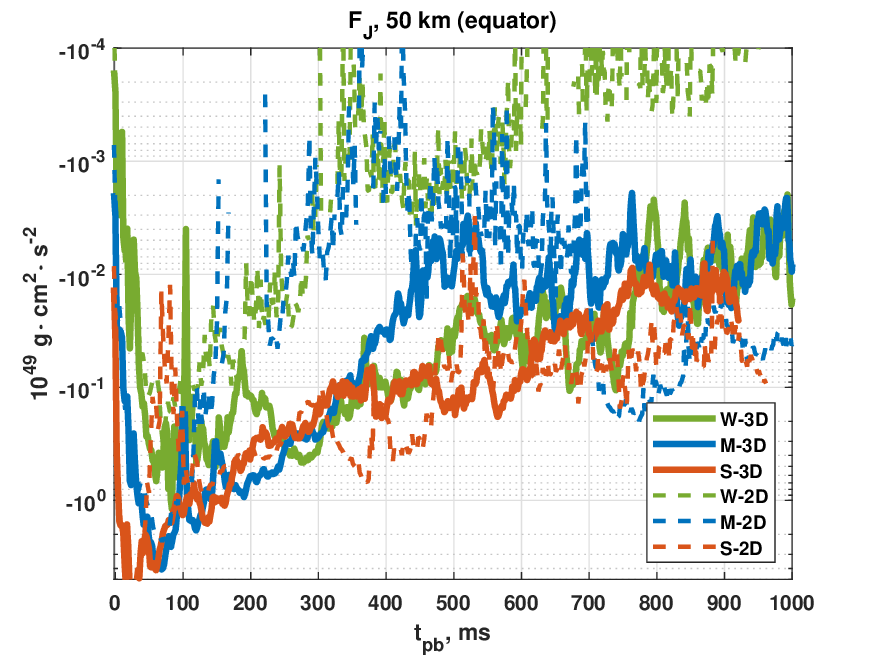}
	\caption{ Evolution of the angular momentum flux from the PNS due to magnetic torques, $F_J$ (Eq.\eqref{torque}), evaluated at the sphere with $r_{PNS}=50$~km: upper panel -- polar flux $F_{J,pol}$; lower panel -- equatorial flux $F_{J,eq}$. Solid thick lines correspond to the 3D models, dashed lines -- to the 2D models. Colors denote the initial field strength: green -- W, blue -- M, red -- S. The fluxes are presented in the units of $F_{J,49} = 10^{49} g\cdot cm^2\cdot s^{-2}$. }
	\label{fig:torque}
\end{figure}

Assuming that the PNS has a radius of $r_{PNS} = 50$ km and neglecting its non-sphericity, the following angular momentum fluxes due to magnetic torques are calculated as 
\begin{equation}
	\begin{gathered}
	F_{J}(\theta_1, \theta_2)  = r_{PNS}^3\int_{0}^{2\pi}d\varphi \int_{\theta_1}^{\theta_2} B_r B_\varphi \sin^2\theta d\theta,
	\end{gathered}
	\label{torque}
\end{equation}
where  angles $\theta_1$ and $\theta_2$ define the regions of polar and equatorial fluxes. These fluxes are correlated with the evolution of $J_{PNS}$. We calculate the equatorial flux as $F_{J,eq} = F_{J}(\pi/3, 2\pi/3)$, and the polar one is evaluated as $F_{J,pol} = F_{J}(0, \pi/3) + F_{J}(2\pi/3, \pi)$. The solid angles for polar and equatorial regions were chosen to be equal to $2\pi$ steradians. The evolution of the angular momentum fluxes is presented in Fig.~\ref{fig:torque} for polar (upper panel) and equatorial (lower panel) regions. 

In the first $\sim 50-80$ ms after the bounce, all six models exhibit a violent outflow of angular momentum that apparently reflects the rapid field wrapping near the bounce, after which the flux magnitude generally decreases. In the polar regions, which are primarily responsible for the jet formation, the hierarchy of models is clear for both 3D and 2D models. Among three-dimensional models, the strong-field model S-3D retains the largest outflow in absolute value, which is naturally associated with the largest  regular part of the magnetic field, that provides an effective magnetic torque. Models M-3D and W-3D exhibit slowly-varying polar outflows of a similar magnitude after $\sim 200$~ms p.b., which partially explains the almost equal explosion energies in these models. M-3D reaches this regime approximately 100 ms after W-3D, which also coincides with a lag in their explosion times, noticeable in Fig.~\ref{fig:Eexpl}. All 3D models show a slow weakening of the magnetic braking, correlated with the overall decrease of the magnetic field. The flux curves for the corresponding 2D models are clearly separated from each other, with the strongest torque in S-2D, followed by M-2D and W-2D. After $t_{pb}\sim 300$~ms, the polar magnetic torques practically reach a plateau with only a noticeable increase in absolute value of the flux in the model M-2D at $t_{pb}\sim 700$~ms. A notable difference from the 3D models is that the flux magnitude here does not show any significant signs of saturation of the angular momentum extraction from the PNS. 

The equatorial region (lower panel of Fig.~\ref{fig:torque}) is more complex and does not exhibit such a clear hierarchy: three-dimensional models form a dense bundle, where individual models are indistinguishable in terms of flux strength. The only common feature among the 3D models is a gradual decrease in the magnetic torque. The W-2D model yields the weakest, near-zero equatorial flux, while M-2D exhibits a noticeable outflow after $t_{pb}\approx 700$~ms. The strong-field model S-2D instead shows a robust outflow throughout the simulated evolution.

Summarizing the presented behavior of the PNS angular momentum evolution in 2D and 3D, one can conclude that 3D models exhibit a more powerful outflow of angular momentum in the polar regions during the first $\sim500$~ms after the core bounce. It leads to a faster conversion of the PNS rotational energy into the explosion compared with the axisymmetric models (see Fig. ~\ref{fig:Eexpl}). Indeed, an increase of the poloidal component of the magnetic field via non-axisymmetric instabilities results in a more effective torque $\sim B_r B_\varphi$, which turns out to be larger than in the corresponding 2D models. At later times, however, the 3D models show a gradual decrease of the momentum outflow due to the magnetic field decay, but the axisymmetric models still convert the PNS differential rotation into magnetic field energy.

\subsubsection{Kink instability}

\begin{figure*}[!htp]
	\centering  
    \includegraphics[width=5.8cm,height=4.7cm]{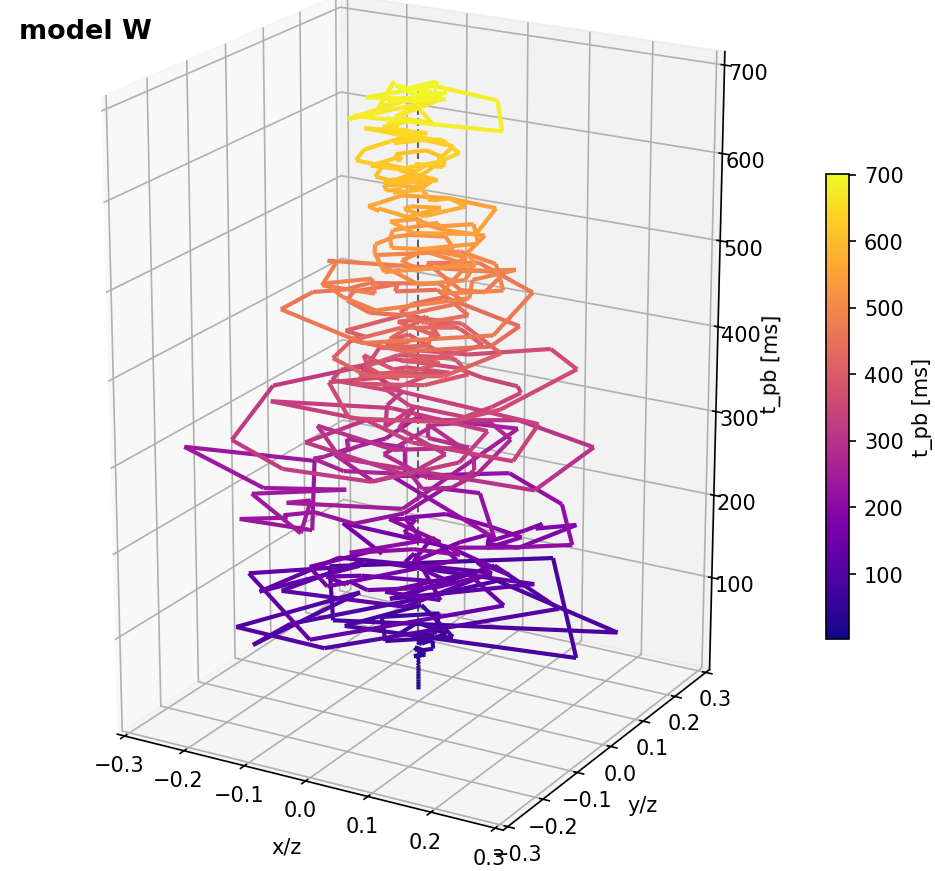}
	\includegraphics[width=5.8cm,height=4.7cm]{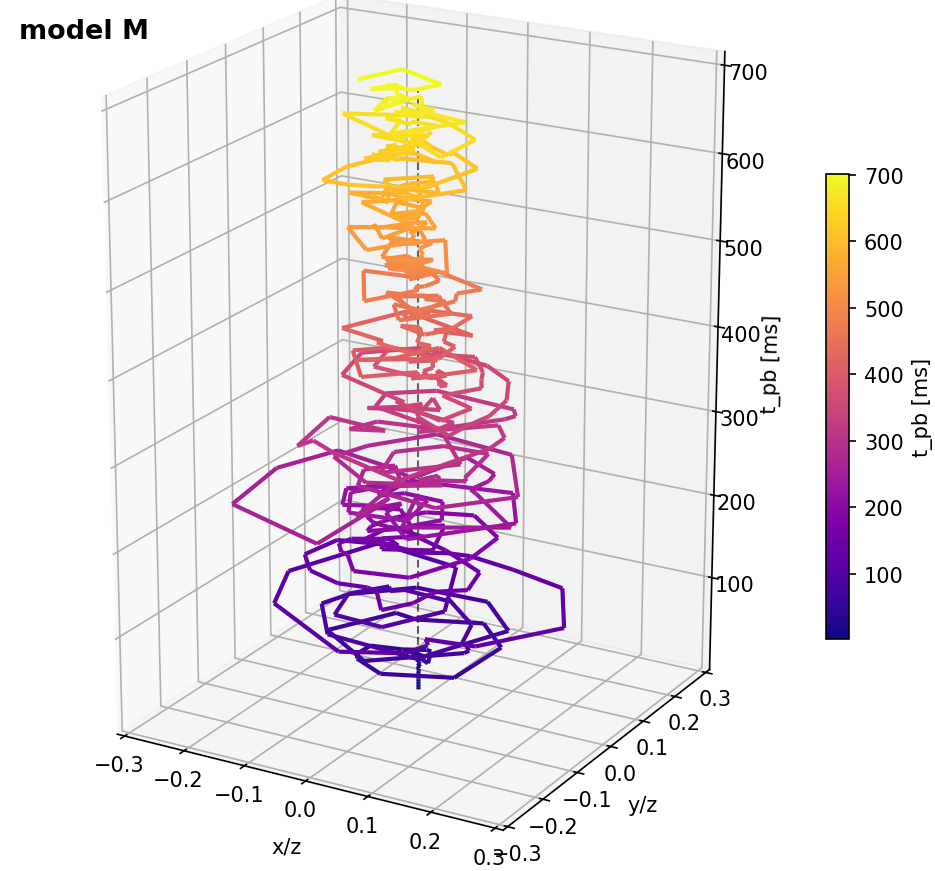}
	\includegraphics[width=5.8cm,height=4.7cm]{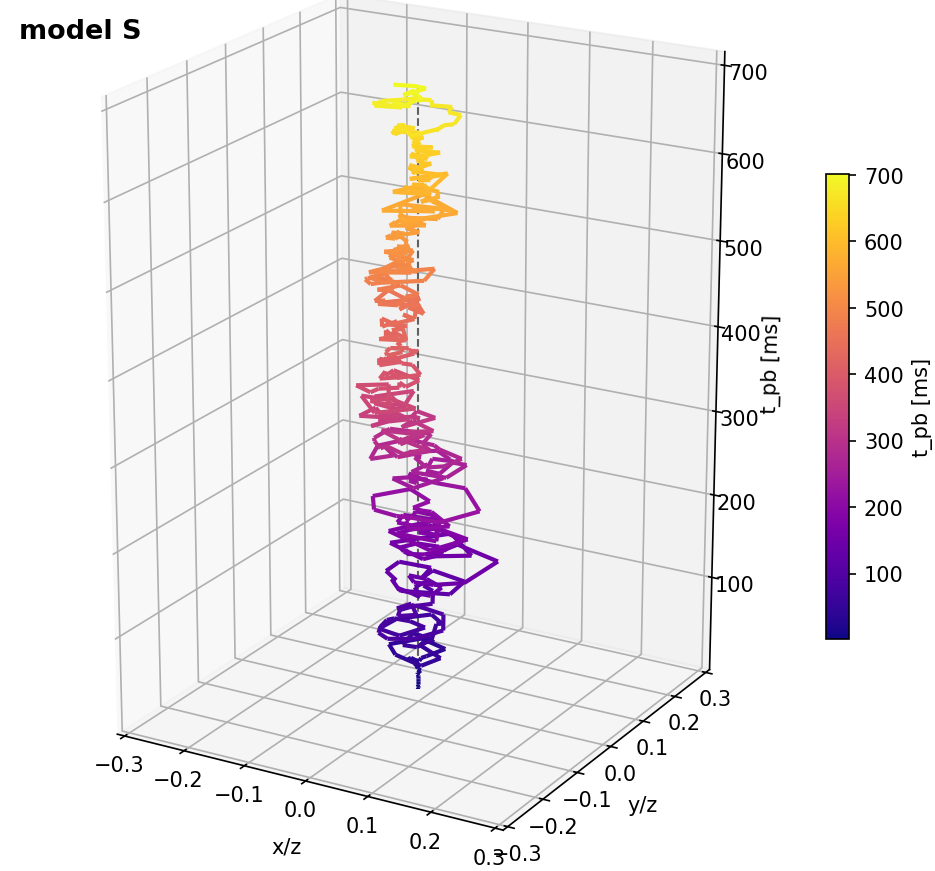}
	\caption{ The evolution of the (normalized) northern jet barycenter position $(X,Y)/|Z|$ with $|Z| = 100$~km for the simulated 3D models: left panel -- W-3D, central panel -- M-3D, right panel --  S-3D.}
	\label{fig:jet_barycenter}
\end{figure*}

In axisymmetric MR models with a large toroidal-to-poloidal field ratio, jet-like outflows are successfully launched. In three dimensions the kink instability is generally considered as the main process responsible for distorting and bending the jets. It is due to more rapid development of a turbulence in 3D models, what could postpone or even destroy the jet collimation \cite{begelman1998, moesta2014}. We have evaluated the effect of the kink instability by tracking the position of the jet barycenter, following the approach used in \cite{moesta2014, kuroda2020, bugli2021, aloy2021, powell2023}. The barycenter position is determined as
\begin{equation}
	\begin{gathered}
	\langle \textit{\textbf{x}}\rangle  = \frac{\int \textit{\textbf{x}} B^2dS}{\int B^2dS},
	\end{gathered}
	\label{jet_barycenter}
\end{equation}
where $\textit{\textbf{x}}=(X,Y)$ are Cartesian coordinates, and the integration is performed over the surface of a horizontal circular disk of radius 100~km located at $Z = 100$~km. Fig.~\ref{fig:jet_barycenter} shows the evolution of the northern jet barycenter at this height (the results for the southern jet are  similar). The plots show that the kink-related displacement at saturation is smaller for larger initial poloidal fields, and that its effect on the outflow morphology decreases with time. The weak- and medium-field models are more affected by the kink instability. The delayed jet formation in the model W-3D, and early outflow evolution in model M-3D are also could be connected with this instability, acting at early post-bounce times. The linear growth phase of the barycenter displacement proceeds over the first tens of milliseconds after the bounce. We have measured e-folding times as $t_{e-fold}\approx4.0$, $3.5$, and $2.8$~ms for the barycenter displacement $\xi = \sqrt{X^2 + Y^2}/|Z|$ in models W-3D, M-3D, and S-3D, respectively. The instability grows faster for stronger initial fields -- despite ultimately saturating at a smaller displacement amplitude. The instability saturates at different times in each model. This time is equal to $\sim80$~ms post-bounce with $\xi\sim0.2$ for W-3D, to $\sim50$~ms with $\xi\sim0.1-0.15$ for M-3D, and to $\sim30$~ms with $\xi\lesssim0.1$ for S-3D. The measured kink instability properties for the strong-field model are of the same order of magnitude as that reported by \cite{bugli2021} for a more massive main sequence progenitor $35\,M_\odot$, rotating at a similar rate.
 
Despite the action of the kink modes, all models exhibit a successful jet break-out. For strong initial fields a minor effect of the kink instability is expected \cite{halevi2018, aloy2021, bugli2021}. The calculations show that the models W-3D and M-3D, having a larger kink-related displacement, also achieve jet launching. 

\section{Summary and discussion} 
\label{sec:summary}

We have presented three-dimensional MHD calculations of the collapse and MR supernova explosion of a rotating magnetized star at the main sequence progenitor mass $20 M_\odot$. The calculations are performed in the frame of the ideal MHD with self-gravity, and approximate neutrino leakage scheme. Three models are considered -- with weak (W-3D, $B_0 = 1/\sqrt{4\pi}\times 10^{12}$~G), medium (M-3D, $B_0 =10^{12}$~G), and strong (S-3D,$B_0 = \sqrt{4\pi}\times 10^{12}$~G) initial poloidal fields. For each 3D model, a corresponding two-dimensional axisymmetric calculation is performed with a similar numerical scheme, the same spacial resolution, and identical initial conditions, and the input physics. This approach allows us to highlight the differences in physical outcomes between 3D and 2D calculations, which are not related to the details of the physical models. 

The main result of the work is the identification, in three-dimensional models with weak and intermediate fields (W-3D, M-3D), of a phase of rapid exponential growth of the poloidal magnetic energy in the PNS with a characteristic growth time of about $10$~ms, which was not found in our corresponding 2D models. The growth is localized mostly in the convectively unstable layer of the PNS. After the termination of this phase, the field acquires a clearly non-axisymmetric turbulent structure, where the ordered component tends out to be subdominant, with the remainder being small-scale fluctuations. By comparing the measured growth rate with the characteristic growth rates of the MRI, the Tayler instability, and convection, we identify a convective instability and the turbulence development, as the main mechanism responsible for the field amplification. The explanation of this conclusion is given in the Section \ref{sec:PNS_mag_field}.

In all three models, the final magnetic field levels converge to comparable magnitudes regardless of the initial field strength, indicating saturation determined by the local properties of turbulence (Fig. \ref{fig:emag_pns}). The amplification of the poloidal magnetic field by non-axisymmetric processes directly affects the explosion dynamics. The increase of the poloidal component results in a more effective magnetic torque $\sim B_r B_\varphi$ and, accordingly, a faster conversion of the PNS rotational energy into the explosion. So, after $\sim200$ ms p.b., the explosion energy in the 3D models increases faster than in their axisymmetric analogues. However, in 3D, this process saturates. From the other side, the turbulent scale fragmentation is less pronounced in 2D what may also lead to almost constant torque -- in the absence of the turbulent scales fragmentation, and the explosion energy continues to increase. In 3D, the explosion always develops faster than in 2D, but it is not necessarily more powerful.
 
Our results on the kink instability are  consistent with the established picture (e.g. \cite{moesta2014, kuroda2020}), where the $m = 1$ mode destroys jet collimation at the  early times of the MR explosion. The measured linear growth times are shorter for higher initial fields, since the toroidal field is stronger there, with saturation achieved at a smaller amplitude for the barycenter displacement compared with the weaker-field models. The kink in our case is indeed active and destructive at early times. In W-3D and M-3D models, there is no collimation of the outflow at the initial stage. The measured kink characteristics (linear growth times $t_{e-fold} \approx 4.0, 3.5, 2.8$ ms and saturation amplitudes $\xi \sim 0.2, 0.1-0.15, \lesssim 0.1$ for the W, M, and S models) are of order of magnitude consistent with the results of Bugli et al. (2021) \cite{bugli2021} for a different progenitor and independent code. Despite the influence of the kink modes, the jets appeared and developed in all simulated models. For weak- and medium-field cases, which are more strongly affected by the convective instability, we attribute the successful launch of the jets to an additional poloidal component generated by non-axisymmetric motions.    

Our obtained convective field amplification connects to a series of idealized semi-global and ``box'' simulations of magnetic field evolution in PNSs \cite{moesta2015, raynaud2020, masada2022, reboulsalze2022, barrere2025}. These works simulate the dynamo in a setting where nondimensional parameters of the flow are under control, but neglect global feedback -- accretion, the evolution of the rotation profile, and the explosion itself. In our global calculations the magnetic field evolution self-consistently develops during collapse and explosion. However, in this approach it is computationally difficult to resolve the small-scale dynamics in detail.  

The magnetic fields can be amplified by the TS cycle which requires a stably stratified medium ($N^2_{BV}>0$) \cite{spruit2002, fuller2019, petitdemange2024}, whereas the growth we observe occurs predominantly in the convectively unstable layer with $N^2_{BV}<0$. In a real PNS, various channels can operate in different zones -- the convection/MRI in the outer layers and the Tayler instability in the stably stratified core, and the overall field amplification pattern likely combines them all.

The obtained results add a self-consistent global argument to the possibility discussed in the literature, that convective field amplification in rapidly rotating PNSs may be a way for generating strong poloidal components even with moderate precollapse magnetic fields, weakening the requirement for unrealistically strong initial magnetic configurations. Our calculations also show that 3D effects change both the efficiency of the MR mechanism and the resulting field topology.

\begin{acknowledgments}
This research was supported in part through computational resources of HPC facilities at HSE University. 
\end{acknowledgments}

\appendix
\subsection*{Appendix: A Simplified Convergence  Study for the Weak-Field Model W-3D}

To evaluate the sufficiency of the spatial numerical resolution used in our simulations, we have performed an additional run of model W-3D on a finer grid, which is denoted as W-3D-f The grid resolution  in W-3D-f is $N_r\times N_\theta\times N_\varphi = 308\times 96 \times 192$. Within the innermost 14~km of the domain we use radial resolution $\Delta r = 0.5$~km, while at larger radii the radial cell size is stretched logarithmically, so that $\Delta r \sim r\,\Delta\theta \sim r\,\sin\theta \,\Delta\varphi$ up to the maximal radius $r_{max}\approx 4.7\times10^{4}$ km. In the post-shock region and in the convectively unstable layer, the resolution of the model  W-3D-f is approximately 1.5 times finer in all three directions than in W-3D. The CPU time used for W-3D-f is $\sim 3$ times larger than for W-3D.

\begin{figure}[!htp]
	\centering
	\includegraphics[width=8.6cm,height=5.7cm]{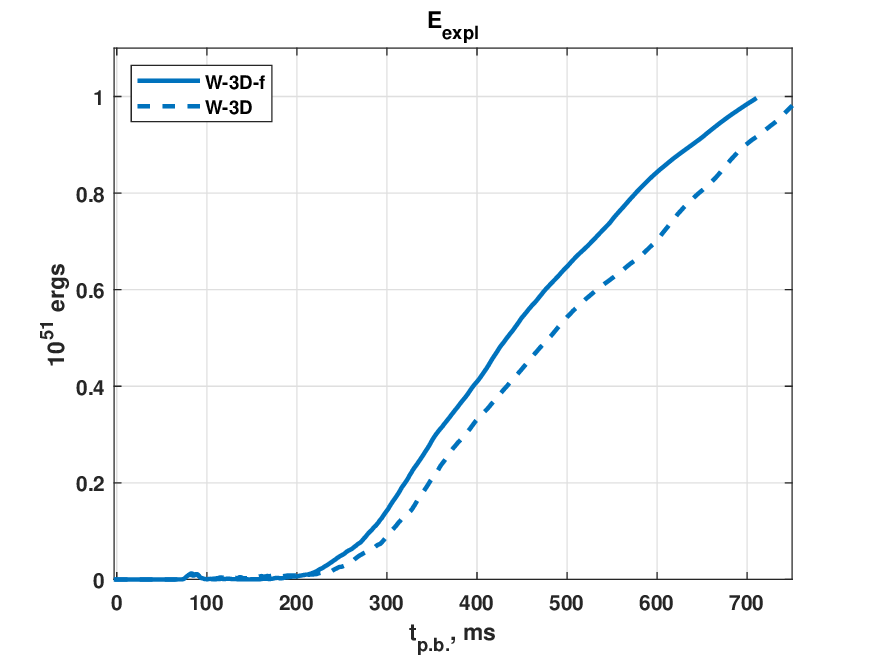}
	\includegraphics[width=8.6cm,height=5.7cm]{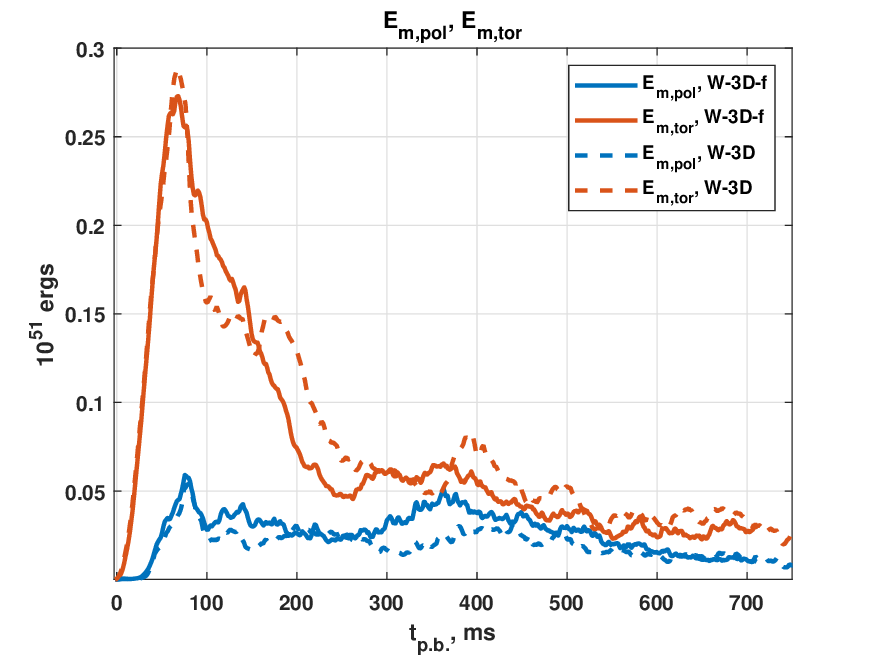}
	\caption{ Integral characteristics for the models W-3D and W-3D-f: upper panel -- explosion energy, lower panel -- poloidal (blue) and toroidal (red) magnetic energies inside the PNS. Solid lines correspond to model W-3D-f, while dashed ones correspond to W-3D.}
	\label{fig:W3D_conv}
\end{figure}

Figure~\ref{fig:W3D_conv} shows the time evolution of the explosion energy and of the magnetic energy inside the PNS for both runs. The plots show that the overall dynamics is essentially the same in both cases: the peaks in PNS magnetic energy occur at the same time and reach nearly identical amplitudes, while the explosion energy is slightly higher for the model W-3D-f. A somewhat higher value of the PNS poloidal magnetic energy in the model W-3D-f is  due to the lower numerical resistivity in the finer-grid.  More effective magnetic torque   in W-3D-f results in $ \sim10\%$ excess of the explosion energy  for the same post-bounce times. The comparison of the W-3D and W-3D-f results shows that the spatial resolution of the numerical grid used in our simulations is acceptable.

\bibliography{kondratyev_3D_MRSN_PRD}
	
\end{document}